\documentclass[usenatbib,onecolumn,doublespacing]{mn2e}

\input epsf
\input psfig.sty

\newcommand{\be}{\begin{equation}}
\newcommand{\ee}{\end{equation}}
\newcommand{\bea}{\begin{eqnarray}}
\newcommand{\eea}{\end{eqnarray}}
\newcommand{\bc}{\begin{center}}
\newcommand{\ec}{\end{center}}

\title[Evolution of planetesimal discs and planets migration]
{Evolution of planetesimal discs and planets migration}
\author[A. Del Popolo, S. Ye\c{s}ilyurt \& N. Ercan]
{A. Del Popolo,$^1$$^,$$^2$$^,$$^3$ S. Ye\c{s}ilyurt$^3$, N. Ercan$^3$\\
$^1$ Dipartimento di Matematica, Universit\`{a} Statale di Bergamo,
  Piazza Rosate, 2 - I 24129 Bergamo, ITALY \\
$^2$ Feza G\"ursey Institute, P.O. Box 6 \c Cengelk\"oy, Istanbul, Turkey\\
$^3$  Bo$\breve{g}azi$\c{c}i University, Physics Department,
 	80815 Bebek, Istanbul, Turkey
}

\begin{document}

\maketitle

\begin{abstract}
In this paper, we further develop the model for the migration of
planets introduced in \citet{DP1} and extended to time-dependent
planetesimal accretion disks in \citet{DP2}. More precisely, the
assumption of \citet{DP2} that the surface density in planetesimals is
proportional to that of gas is released. Indeed, the evolution of the
radial distribution of solids is governed by many processes: gas-solid
coupling, coagulation, sedimentation, evaporation/condensation, so
that the distribution of planetesimals emerging from a turbulent disk
does not necessarily reflect that of gas (e.g., \citealt{SV1}). In
order to describe this evolution we use a method developed in
\citet{SV2} which, using a series of simplifying assumptions, is able
to simultaneously follow the evolution of gas and solid particles for
up to $10^7 {\rm yr}$. This model is based on the premise that the
transformation of solids from dust to planetesimals occurs through
hierarchical coagulation. Then, the distribution of planetesimals
obtained after $10^7 {\rm yr}$ is used to study the migration rate of
a giant planet through the migration model introduced in
\cite{DP1}. This allows us to investigate the dependence of the
migration rate on the disk mass, on its time evolution and on the
value of the dimensionless viscosity parameter $\alpha$. We find that
in the case of disks having a total mass of $10^{-3}-10^{-1}
M_{\odot}$, and $10^{-4}<\alpha<10^{-1}$, planets can migrate inward
over a large distance while if $M_{\rm d}<10^{-3} M_{\odot}$ the planets
remain almost at their initial position for $\alpha>10^{-3}$ and only
in the case $\alpha<10^{-3}$ the planets move to a minimum value of
orbital radius of $\simeq 2 {\rm AU}$. Moreover, the observed
distribution of planets in the period range 0-20 days can be easily
obtained from our model. Therefore, dynamical friction between planets
and the planetesimal disk provides a good mechanism to explain the
properties of observed extra-solar giant planets.
\end{abstract}


\section{Introduction}

The discovery of solar-like stars showing evidences for planets
orbiting around them (\citealt{Mayor1}, \citealt{Marcy1},
\citealt{Vogt1}, \citealt{Butler1}) has greatly intensified the
interest in understanding the formation and evolution of planetary
systems, as well as the long-standing problem of the solar system
origin. At the same time these discoveries have raised a number of
questions about the formation mechanisms of such systems. Indeed, the
extra-solar planets discovered so far are all more massive than
Saturn, and most either orbit very close to their stars or travel on
much more eccentric paths than any of the major planets in our Solar
System.  The present small sample can be broadly divided into three
groups:\\
a) Jupiter analogues, in terms of period $P$ and semi-axis $a$, and with low 
eccentricities (such as 47~UMa);\\
b) planets with highly eccentric orbits (such as 70~Vir); \\
c) close-in giants (or `hot Jupiters') within 0.1~AU whose orbits are largely 
circular (such as 51~Peg). \\

The properties of these planets, most of which are Jupiter-mass objects,
are difficult to explain using the standard model for planet formation 
(\citealt{Lissauer1}; \citealt{Boss1}). Current theories (\citealt{Mizuno1}; 
\citealt{Bod1}) predict that giant planets were formed by gas accretion onto 
massive ($ \simeq 15 M_{\rm \oplus}$) rocky cores which themselves were the 
result of the accumulation of a large number of icy planetesimals. The most 
favorable conditions for this process are found beyond the so-called 
``snow line'' (\citealt{Hayashi1}; \citealt{Sasselov1}). As a consequence, 
this standard model predicts nearly circular planetary orbits and giant
planets distances $\geq 1$ AU from the central star where the temperature 
in the protostellar nebula is low enough for icy materials to condense 
(\citealt{Boss1}; \citealt{Boss2}; but see also \citealt{Wuchterl1};
\citealt{Wuchterl2}). 
%
%%Indeed, standard disk models show that at 0.05 AU (region
%%in which were found some of the planets), the temperature is about 2,000 K, 
%%which is too hot for the existence of any small solid particles.
%
Thus, in the case of close-in giants, it is very unlikely that such planets 
were formed at their present locations. Then, the most 
natural explanation for this paradox and for planets on very short orbits is
that these planets have formed further away in the protoplanetary
nebula and they have migrated afterwards to the small orbital distances at
which they are observed. Some authors have also proposed scenarios
in which migration and formation were concurrent \citep{Terquem1}. 
We refer the reader to \citealt{DP1} (hereafter DP1) and \citealt{DP2} 
(hereafter DP2) for a detailed discussion of the different mechanisms that 
have been proposed to explain the presence of planets at small orbital 
distances: (a) dynamical instabilities in a 
system of giant planets (\citealt{Rasio1}; \citealt{Weid1}), 
(b) ``migration instability'' \citep{Murray1}, (c) tidal interaction with 
a gaseous nebula (\citealt{Gold1}; \citealt{Gold2}; \citealt{Ward1}; 
\citealt{Lin4}; \citealt{Ward2}) and (d) dynamical friction between the 
planet and a planetesimal disk (DP1).

In particular, in DP1 and DP2 we showed that dynamical friction
between a planet and a planetesimals disk is an important mechanism
for planet migration and we pointed out that some advantages of the
model are:
\\ 
a) Planet halt is naturally provided by the model. \\ 
b) It can explain planets found at heliocentric distances of
$>0.03-0.04$ AU, or planets having larger values of eccentricity. \\
c) It can explain metallicity enhancements observed in stars having
planets in short-period orbits.\\
d) Radial migration is possible with modest masses of planetesimal
disks, in contrast with other models.

For the planetesimal disk used in DP1, following \citet{Opik1}, we
assumed that the surface density in planetesimals $\Sigma_{\rm s}$
varies as $\Sigma_{\rm s}(r)=~\Sigma_{\odot}(1 {\rm AU}/r)^{3/2}$,
where $\Sigma_{\odot}$, the surface density at 1 AU, was a free
parameter. In DP2 the previous assumption was substituted by a more
reliable model for the disk, and in particular we used a
time-dependent accretion disk, since it is widely accepted that the
solar system at early phases in its evolution is well described by
this kind of structure. An important assumption of DP2 was that the
surface density in planetesimals remains proportional to that of gas:
$\Sigma_{\rm s}(r,t) \propto \Sigma(r,t)$. However, it is well-known
that the distribution of planetesimals emerging from a turbulent disk
does not necessarily reflect that of gas (e.g., \citealt{SV1},
\citealt{SV2}). Indeed, in addition to gas-solid coupling, the
evolution of the distribution of solids is also governed by
coagulation, sedimentation and evaporation/condensation. In order to
take into account these effects we use the method developed in
\citet{SV2} which is able to simultaneously follow the evolution of
gas and solid particles for up to $10^7 {\rm yr}$. The main
approximation used in this model is to associate one grain size to a
given radius and time. Then, we use the radial distribution of
planetesimals predicted by this model to estimate the planet
migration, which is computed as in \citet{DP1}.

This paper is organized as follows. In Sect.\ref{Gas and solids
distribution and evolution} we describe the disk model we use to
obtain the radial distribution of the planetesimal disk reached after
$10^7$ yr. Then, in Sect.\ref{Migration model} we briefly review the
migration model introduced in \citet{DP1}. Finally, we describe our
results in Sect.\ref{Results}.

\section{Disk model and planet migration}

\subsection{Gas and solids distribution and evolution}
\label{Gas and solids distribution and evolution}

It is well-known that protostellar disks around young stellar objects
that have properties similar to that expected for the solar nebula are
common: between 25\% to 75\% of young stellar objects in the Orion
nebula seem to have disks with mass $10^{-3} M_{\odot}<M_{\rm d}
<10^{-1} M_{\odot}$ and size $40 \pm 20$ AU
\citep{Beckwith1}. Moreover, observations of circumstellar disks
surrounding T Tauri stars support the view of disks having a limited
life-span and characterized by continuous changes during their
life. These evidences have led to a large consensus about the nebular
origin of the Solar System. Moreover, it clearly appears necessary to
model both the spatial and temporal changes of the disk (which cannot
be handled by the minimum-mass model nor by steady-state
models). Besides, one also needs to describe the global evolution of
the solid material which constitutes, together with the gas, the
protoplanetary disk. Of particular interest is the spatial
distribution of material making up a planetary system, as this is
about the only information the present observations, based on the
Doppler technique, can provide. The knowledge of this distribution and
its time evolution is important to understand how planets form and in
this paper it is a key issue since we wish to study the
planet migration due to the interaction between planets and the local
distribution of solid matter.

As usual, the time evolution of the surface density of the gas
$\Sigma$ is given by the familiar equation (e.g., \citealt{SV2}):
\be
\frac{\partial \Sigma}{\partial t} -\frac{3}{r}\frac{\partial
}{\partial r}\left[ r^{1/2}\frac{\partial }{\partial r}\left(
r^{1/2}\nu _{t}\Sigma \right) \right] =0
\label{gasevol}
\ee
where $\nu_{\rm t}$ is the turbulent viscosity. Since $\nu_{\rm t}$ is
not an explicit function of time, but instead depends only on the
local disk quantities, it can be expressed as $\nu_{\rm t}=\nu_{\rm
t}(\Sigma,r)$ and Eq.(\ref{gasevol}) can be solved subject to
boundary conditions on the inner and outer edges of the disk. The
opacity law needed to compute $\nu_{\rm t}$ is obtained from
\citet{Ruden1}. Then, Eq.(\ref{gasevol}) is solved by means of an
implicit scheme. Note that the evolution of the gas is computed
independently from the evolution of particles (which only make $\sim
1\%$ of the gas mass). Next, from $\Sigma(r,t)$ we can algebraically
find all other gas disk variables.

Next, as described in \citet{SV2} the evolution of the surface density
of solid particles $\Sigma_{\rm s}$ is given by:
\be
\frac{\partial \Sigma_{\rm s} }{\partial t} = \frac{3}{r} \frac{\partial
}{\partial r} \left[ r^{1/2} \frac{\partial}{\partial r} (\nu_{\rm s}
\Sigma_{\rm s} r^{1/2}) \right] + \frac{1}{r}\frac{\partial }{\partial
r} \left[ \frac{2r \Sigma _{s} \langle \overline{v}_{\phi}
\rangle_{s}} {\Omega _{k}t_{s}} \right] .
\label{solidevol}
\ee
The first diffusive term is similar to Eq.(\ref{gasevol}), where the
effective viscosity $\nu_{\rm s}$ is given by:
\be
\nu_{\rm s} = \frac{\nu_{\rm t}}{{\rm Sc}} \hspace{0.4cm} \mbox{with}
\hspace{0.4cm} {\rm Sc} = \left( 1+\Omega _{\rm k} t_{\rm s} \right)
\sqrt{1+\frac{\overline{\bf v}^2}{V_{\rm t}^2}} .
\label{Schmidt1}
\ee
Here we introduced the Schmidt number Sc which measures the coupling
of the dust to the gas turbulence. We also used the relative
velocity ${\bf v}$ between a particle and the gas, the turbulent
velocity $V_{\rm t}$, the Keplerian angular velocity $\Omega _{\rm k}$
and the so-called stopping time $t_{\rm s}$. The dimensionless
quantity $(\Omega _{\rm k}t_{\rm s})$ measures the coupling of the
solid particles to the gas. Thus, small particles (size $s<1$ mm) have
$\Omega _{\rm k}t_{\rm s} \ll 1$ and they are strongly coupled to
gas. Therefore, they exhibit the same radial velocity. On the other
hand, large particles ($s>10^4$ cm) with $\Omega _{\rm k}t_{\rm s} \gg
1$ show a much smaller radial drift. Finally, particles in the
intermediate regime ($s \sim 10$ cm) with $\Omega _{\rm k}t_{\rm s} \sim
1$ exhibit a large inward radial velocity. Therefore, the evolution of the
dust radial distribution can be significantly different from the
behaviour of the gas, depending on the particle size (see
\citealt{SV1} for a detailed study).  

The second advective term in Eq.(\ref{solidevol}) arises from the lack
of pressure support for the dust disk as compared with the gas
disk. Thus, it is proportional to the azimuthal velocity difference
$\overline{v}_{\phi}$ between the dust and the gas. The average
$\langle .. \rangle_{s}$ refers to the vertical averaging over the
disk height weighted by the solid density. We refer the reader to
\citet{SV2} for a more detailed presentation, see also \citet{Kornet1}.

In addition to the radial motion described by Eq.(\ref{solidevol}),
the dust surface density also evolves through evaporation/condensation
and coagulation. In this article, following \citet{SV2} we assume that
the size distribution of solid particles at a given orbital radius and
time is narrowly peaked around a mean value $s(r,t)$. This is
supported by numerical simulations \citep{Mizuno2} which show that
although a broad size distribution is maintained, most of the mass is
always concentrated in the largest particles so that one can define a
meaningful typical size $s(r,t)$. Then, within this approximation
coagulation does not influence the dust surface density $\Sigma_{\rm
s}$ since it conserves the total mass of solids. Thus, the coagulation
of solid particles only appears through the evolution of the radial
distribution $s(r,t)$ of the typical size of the dust grains. We
model this process as in \citet{SV2}. We must note that we only
consider collisional coagulation and we disregard gravitational
interactions which would come into play at late times when large
planetesimals have formed.

On the other hand, we take into account the evaporation of solid
particles which takes place at the radius $r_{\rm evap}$ where the
temperature reaches $T_{\rm evap}$. We also include the condensation
of the vapor below $T_{\rm evap}$ onto the solid grains. The velocity
of the vapor is equal to gas velocity but this component evolves in a
specific way because of its own diffusion process and condensation. In
this article we are mainly interested in the distribution of solids at
small radii hence we consider only one species of solid particles:
high-temperature silicates with $T_{\rm evap}= 1350$ K and a bulk
density $\rho_{\rm bulk} = 3.3$ g cm$^{-3}$. Thus, in our simplified
model we follow the evolution of three distinct fluids: the gas, the
vapour of silicates and the solid particles.

In this fashion, we obtain the radial distribution of the planetesimal
swarm after $10^7$ yr. This yields the surface density of solids
$\Sigma_{\rm s}(r,t)$ and the mid-plane solid density $\rho_{\rm
s}(r,t)$. We also obtain the size distribution $s(r,t)$ reached by
hierarchical coagulation. Of course, at these late times where
planetesimals have typically reached a size of a few km or larger,
gravitational interactions should play a dominant role with respect to
coagulation. However, if these interactions do not significantly
affect the radial distribution of solids (note that the radial
velocity of such large particles due to the interaction with the gas
is negligible) we can still use the outcome of the fluid approach
described above to study the migration of giant planets, as detailed
below.

\subsection{Migration model}
\label{Migration model}

In order to study the migration of giant planets we use the model
developed in DP1 (see also DP2). Since this model has already been
described in these two papers, we only recall here the main points. We
consider a planet revolving around a star of mass $M_{\ast}=1
M_{\odot}$. As described in the introduction, in this paper we suppose that the 
formation mechanism is core accretion and we are then not interested by the scenarios 
that gravitational instability models could introduce in our model.

The equation of motion of the planet can be written as:
\be
{\bf \ddot r}= {\bf F}_{\odot} + {\bf R}
\ee
where the term ${\bf F}_{\odot}$ represents the force per unit mass
from the star, while ${\bf R}$ is the dissipative force (i.e. the
dynamical friction term--see \citealt{Melita1}). If we assume a
disk-shaped matter distribution with constant velocity dispersions
$\sigma_{\parallel}$ (parallel to the plane) and $\sigma_{\perp}$
(perpendicular to the plane) and with a ratio simply taken to be 2:1
(i.e. $\sigma_{\parallel}$=$2 \sigma_{\perp}$), then according to
\citet{Chandra1} and \citet{Binney1} we may write the force components as:
\[
F_{\parallel} = k_{\parallel}v_{1 \parallel}
\]
\be
\;\; = B_{\parallel}v_{1\parallel } \left[ 2\sqrt{2\pi } \overline{n}
G^2\log \Lambda m_1m_2\left( m_1+m_2\right) \frac
{\sqrt{1-e^2}}{\sigma _{\parallel }^2\sigma _{\perp }}\right] ,
\label{eq:b1}
\ee
\[
F_{\perp } = k_{\perp}v_{1 \perp}
\]
\be
\;\; = B_{\perp }v_{1_{\perp }} \left[ 2\sqrt{2\pi } \overline{n}
G^2\log \Lambda m_1m_2\left( m_1+m_2\right) \frac
{\sqrt{1-e^2}}{\sigma _{\parallel }^2\sigma _{\perp }}\right] ,
\label{eq:b2}
\ee
where
\begin{eqnarray}
B_{\parallel} & = &  \int_0^\infty \frac {dq}{(1+q)^2 (1-e^2+q)^{1/2}}
\nonumber \\ 
& & \times \exp{ \left
[ -\frac{v_{1\parallel}^2}{2\sigma_{\parallel}^2} \frac{1}{1+q} -
\frac{v_{1\perp}^2}{2\sigma _{\parallel}^2} \frac {1}{1-e^2+q} \right]
} ,
\label{eq:b3}
\end{eqnarray}
\begin{eqnarray}
B_{\perp } & =  & \int_0^\infty \frac{dq}{(1+q) (1-e^2+q)^{3/2}}
\nonumber \\ 
& & \times \exp{ \left[ -\frac{v_{1\parallel}^2}{2\sigma
_{\parallel}^2} \frac{1}{1+q} - \frac{v_{1\perp}^2}{2\sigma
_{\parallel}^2} \frac{1}{1-e^2+q} \right] } ,
\label{eq:b4}
\end{eqnarray}
and
\be
e=(1-\sigma_{\perp}^2/\sigma_{\parallel}^2)^{0.5} .
\ee
Here $\overline{n}$ is the average spatial density of field particles,
$m_1$ is the mass of the test particle, $m_2$ is the mass of a field
one, and $\log{\Lambda}$ is the Coulomb logarithm. Then, the
frictional drag on the test particles may be written as:
\be
{\bf F}=-k_{\parallel}v_{1 \parallel} {\bf e_{\parallel}}-
k_{\perp}v_{1 \perp}{\bf e_{\perp}} \label{eq:dyn}
\ee
where ${\bf e_{\parallel}}$ and ${\bf e_{\perp}}$ are two unit
vectors parallel and perpendicular to the disk plane.

Since the damping of eccentricity and inclination is more rapid than
radial migration (\citealt{Ida1}; \citealt{Ida2}; DP1), we only deal
with radial migration and we assume that the planet has negligible
inclination and eccentricity ($i_{\rm p} \sim e_{\rm p} \sim 0$) and
that the initial distance to the star of the planet is $5.2$ ${\rm
AU}$. \footnote{The planet was set at an initial distance of 5.2 ${\rm
AU}$ for similitudes with choices done in previous papers (e.g., Murray et al. 1998, Trilling et al. 1998)} 
For the objects lying in the plane, the dynamical drag is
directed in the direction opposite to the motion of the particle and
is given by:
\be
{\bf F} \simeq -k_{\parallel}v_{\parallel} {\bf e_{\parallel}} .
\ee

In order to calculate the effect of dynamical friction on the orbital 
evolution of the planet, we suppose that
$\sigma_{\parallel}$=$2 \sigma_{\perp}$ and that the dispersion 
velocities
are constant. If the planetesimals attain dynamical equilibrium, their
equilibrium velocity dispersion, $\sigma_{\rm m}$, would be comparable 
to
the surface escape velocity of the dominant bodies (Safronov 1969) such 
that
\begin{equation}
\sigma_{\rm m} 
%\sim v_{\rm esc} 
\sim \left(\frac{G m_{\ast} }{ \theta r_{\ast}}\right)^{1/2}
%
%%\sigma_{\rm m} \sim v_{\rm esc} \sim \left(\frac{2G M_{\rm p} }{R_{\rm
%%p}}\right)^{1/2}
%
%\sim 0.1 \left(\frac{M_{\rm f}}{10^{23}{\rm g}}\right)^{1/3}
%\left(\frac{\rho_{\rm f}}{1{\rm g cm}^{-3}}\right)^{1/2}
%{\rm km/s}.
\label{eq:vesc}
\end{equation}
where $\theta$ is the Safronov number, 
%%$2<\theta<5$,
$m_{\ast}$ and $r_{\ast}$ are the mass and radius of the largest
planetesimals, 
(note that the planetesimals velocity dispersion, $\sigma_{\rm m}$,
now introduced, is the velocity dispersion to be used for
calculating the $\sigma$ which is present in the dynamical friction 
force).
If instead we consider a two-component system, consisting of one 
protoplanet and many equal-mass planetesimals 
%(we remember that planetesimals with a
%general mass distribution are well described by equal mass planetesimals
%with an effective mass $m_{\rm eff}$ (Ida \& Makino 1993),
the velocity dispersion of planetesimals in the neighborhood of the
protoplanet depends on the mass of the protoplanet. When the mass of the
planet, $M$, is $\le 10^{25}$ g, the value of $<e^2_{\rm m}>^{1/2}$
(being $e_{\rm m}$ the eccentricity of the planetesimals) is
independent of $M$ therefore:
\begin{equation}
e_{\rm m} \simeq 20 (2 m /3 M_{\odot})^{1/3}
\end{equation}
(Ida \& Makino 1993) where $m$ is the mass of the planetesimals. When 
the
mass of the planet reaches values larger than $10^{25}$-$10^{26}$ g at
1 AU, $<e^2_{\rm m}>^{1/2}$ is proportional to $M^{1/3}$:
\begin{equation}
e_{\rm m} \simeq 6 (M/3M_{\odot})^{1/3}
\end{equation}
(Ida \& Makino 1993).
As a consequence also the dispersion velocity in the disc is
characterized by two regimes being it connected to the eccentricity
by the equation:
\begin{equation}
\sigma_{\rm m} \simeq (e_{\rm m}^2+i_{\rm m}^2)^{1/2} v_{\rm c}
\end{equation}
where $i_{\rm m}$ is the inclination of planetesimals and
$v_{\rm c}$ is the Keplerian circular velocity.
Following \citet{Stern1} and 
\citet{DSG} we assume that $\langle i_{\rm m}^2 \rangle = \langle
e_{\rm m}^2 \rangle/4$. 
In the simulation we assume that the planetesimals have all 
equal
masses, $m$, and that $m<< M$, $M$ being the planet mass.
This assumption does not affect the results, since dynamical friction does not depend
on the individual masses of these particles but on their overall 
density.
Note that for $m \ll M$ the
frictional drag $F_{\parallel}$ obtained in Eq.(\ref{eq:b1}) does not
depend on the mass $m$ of the planetesimals since the velocity
dispersion $\sigma_{\rm m}$ only depends on the mass $M$ of the giant
planet while the explicit dependence on $m$ of Eq.(\ref{eq:b1})
only involves the product $\overline{n} m = \rho_{\rm s}$. Therefore, we do
not need to follow the evolution of the size distribution of
planetesimals. We merely use the planetesimal density $\rho_{\rm s}$
reached after $10^7$ yr, assuming that the height of the planetesimal
disk does not evolve significantly.

An important point to discuss here is the back reaction of the planet on the swarm.  
As previously reported, in the calculation of the scattering of planetesimals by a protoplanet, 
Ida \& Makino (1993) showed, by means of N-body simulations, that eccentricities, $e_{\rm m}$, inclinations, 
$i_{\rm m}$ and velocity dispersions, $\sigma_{\rm m}$, of planetesimals in the vicinity of the protoplanet
are strongly influenced by the mass of the protoplanet. In the early stage, random velocities of small 
planetesimals remain low during the growth of the protoplanet, since they are regulated by gravitational scatterings between planetesimals. When the protoplanet becomes massive enough ($10^{25}-10^{26}$ g) to influence the velocity distribution of small planetesimals, the random velocities, and velocity dispersion of planetesimals are heated by the protoplanet and become larger than in the early stage. Furthermore, the protoplanet would scatter neighboring planetesimals and give rise to a gap in the spatial distribution of planetesimals (see Fig. 3 of Ida \& Makino 1993, and Fig. 1 of Tanaka \& Ida 1997). 
As noticed by Rafikov (2001) the gaps seen in N-body simulations are never clean because random motion of planetesimals is naturally included, and this permits some of them to be present in the gap.
The process of clearing a gap in a planetesimal disc around a massive body is analogous to gap formation in gaseous discs (Takeuchi et al. 1996; Rafikov 2001).
The gap also reduce the growth rate of the protoplanet. 
In our calculation, this effect was taken into account, as previously reported. 
The width of the heated region is roughly given by
$4 [(4/3)(e_{\rm m}^2+i_{\rm m}^2)a^2+12 h_{\rm M}^2 a^2]^{1/2}$ (Ida \& 
Makino 1993)
where $a$ is the semi-major
axis and $h_{\rm M}= (\frac{M+m}{3 M_{\odot}})^{1/3}$
is the Hill radius of the protoplanet.
The effect on the drift velocities can be easily predicted observing that 
the increase in velocity dispersion of planetesimals around the protoplanet decreases the
dynamical friction force (see Eq. \ref{eq:dyn}) and consequently 
increases the migration time-scale. In order to perform an order of magnitude estimation, we suppose that 
the planet moves on a stable circular orbit. 
%with velocity $v_{\rm c}=\sqrt{G M_\odot/r}$. 
Using for simplicity Chandrasekhar (1943) formula, the drift velocity can be expressed as:
\begin{equation}
\frac{d r}{dt} \simeq -\frac{4 \pi \log{\Lambda} G^2 M \rho(r)r}{\sigma^3}
%{v_{\rm c}^3}
\end{equation}
(Chandrasekhar 1943, Binney \& Tremaine 1987, Palmer et al. 1993, Hernandez \& Gilmore 1998).
Recalling the mass dependence of the velocity dispersion in the two regimes, we find that the drift
velocity behaves as:
\begin{equation}
\frac{d r}{dt} \propto \left\{
\begin{array}{lc}
M & M \le 10^{25} g  \\
constant & M > 10^{25} g
\end{array}
\right.
\end{equation}
%Since the mass of our protoplanet is $\propto M^{1/3}$......mostrare la dipendenza lineare prima e costante dopo..... 
The linear dependence on the planet's mass, of the drift velocity, corresponds to the $type$ I drift in the density wave approach (Ward 1997), while the part of the plot independent on the planet's mass corresponds to $type$ II drift. 
The transition between the two regimes entails a velocity drop of between one to three orders of magnitude (according to the value of $\alpha$).
What previously described is shown in Fig. 1, where we show the drift velocity, $\frac{d r}{d t}$, as function
of mass, $M$, of the protoplanet for $M_{\rm d}=0.1$ and $\alpha=10^{-4}$. As shown, objects having masses $ < M_{\oplus}$ have 
velocity drift increasing as $M$, while after a threshold mass any 
further mass increases begins to slow down the drift. As the threshold 
is exceeded
the motion fairly abruptly converts to a slower mode in which the
drift velocity is independent of mass. As previously explained, this
behavior is due to the transition from a stage in which the dispersion
velocity is independent of $M$ to a stage in which it increases with 
$M^{1/3}$ (Ida \& Makino 1993). This last stage is known as the 
protoplanet-dominated stage.
The phenomenon is equivalent to that predicted in the density wave 
approach
(Goldreich \& Tremaine 1980; Ward 1997). In this approach, the density 
wave torques
repel material on either side of the protoplanet's orbit and attempt to
open a gap in the disc. Only very large objects are able to open and
sustain the gap. After gap formation, the drift rate of the planet is
set by disc viscosity and is generally smaller than in absence of the 
gap.
We stress that the decaying portion of the curve
corresponding to
the transition from the first to the second stage does not correspond to
any particular model because following Ida \& Makino (1993) we do not 
have information on the evolution of $\sigma$ in the transition regime.
%% We also stress that the behaviour $\propto M$ is valid in the $type$ I
%% regime for $M <0.1 M_{\odot}$. 
Even if not necessary, in order to have an independent check of the effects of back reaction, we performed another calculation. 
Suppose to have the following two situations:\\
a) there is an initial gap in the disc;\\
b) the planet is initially embedded in an unperturbed disc and must clear material in order to form a gap.\\
According to Nelson et al. (1999) the quoted situations lead to slightly different results in migration, results that tend to converge with increasing time (see Fig. 3 of Nelson et al. 1999). The small initial difference in migration rate is due to the fact that in situation b 
%the disc is initially more massive in the neighborhood of the planet???????????????, then the migration is more rapid. 
the clearing of the disc material leads to a period of more rapid migration.
Using this result, we compared the migration obtained assuming:\\
1) an initial gap as that shown in Fig. 9 of Nelson et al. (1999) and with $\sigma={\rm constant}$ (not given by Ida \& Makino (1993) relation) 
with\\
2) the migration obtained taking account of the velocity dispersion dependence above described. \\
The results were in good agreement.  

\begin{figure}
\psfig{file=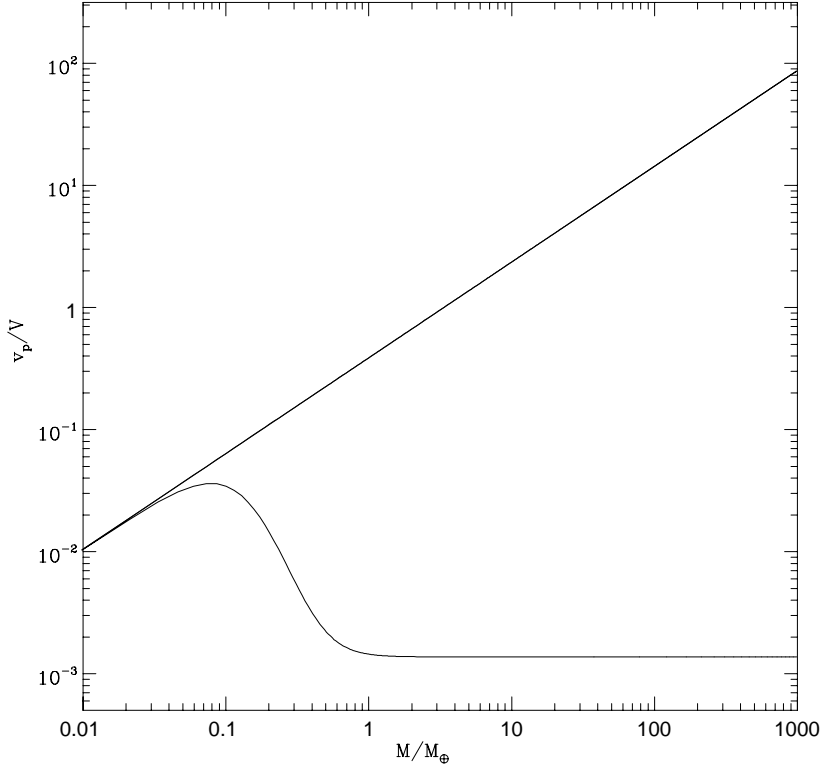,width=12cm}
\caption[]{Drift velocity, $\frac{d r}{d t}$, as a function of mass.
Velocities are normalized to $V=2 \frac{M_{\oplus}}{M_{\odot}}
\frac{\pi \Sigma r^2}{M_{\odot}} (\frac{r \Omega}{\sigma})^3 r \Omega$
where $M_{\oplus}$ is an Earth mass, $\Sigma$ the surface density, 
$\Omega$
is the angular velocity and $\sigma$ the dispersion velocity. The 
assumed
conditions are those considered appropriate for the Jovian region and
assuming that $M_{\rm D}=0.1 M_{\odot}$, $\alpha=10^{-4}$. The line $\propto M$ 
correspond to $type$ I drift 
described by Ward (1997)
and its behavior is valid till
$\simeq 0.1 M_{\oplus}$ but beyond this value there is a transition to
a behavior $\propto M^0$ ($type$ II motion)}.
\label{Fig1}
\end{figure}

\section{Results}
\label{Results}

\subsection{General framework}

\begin{figure*}
\centerline{\hbox{(a)
\psfig{figure=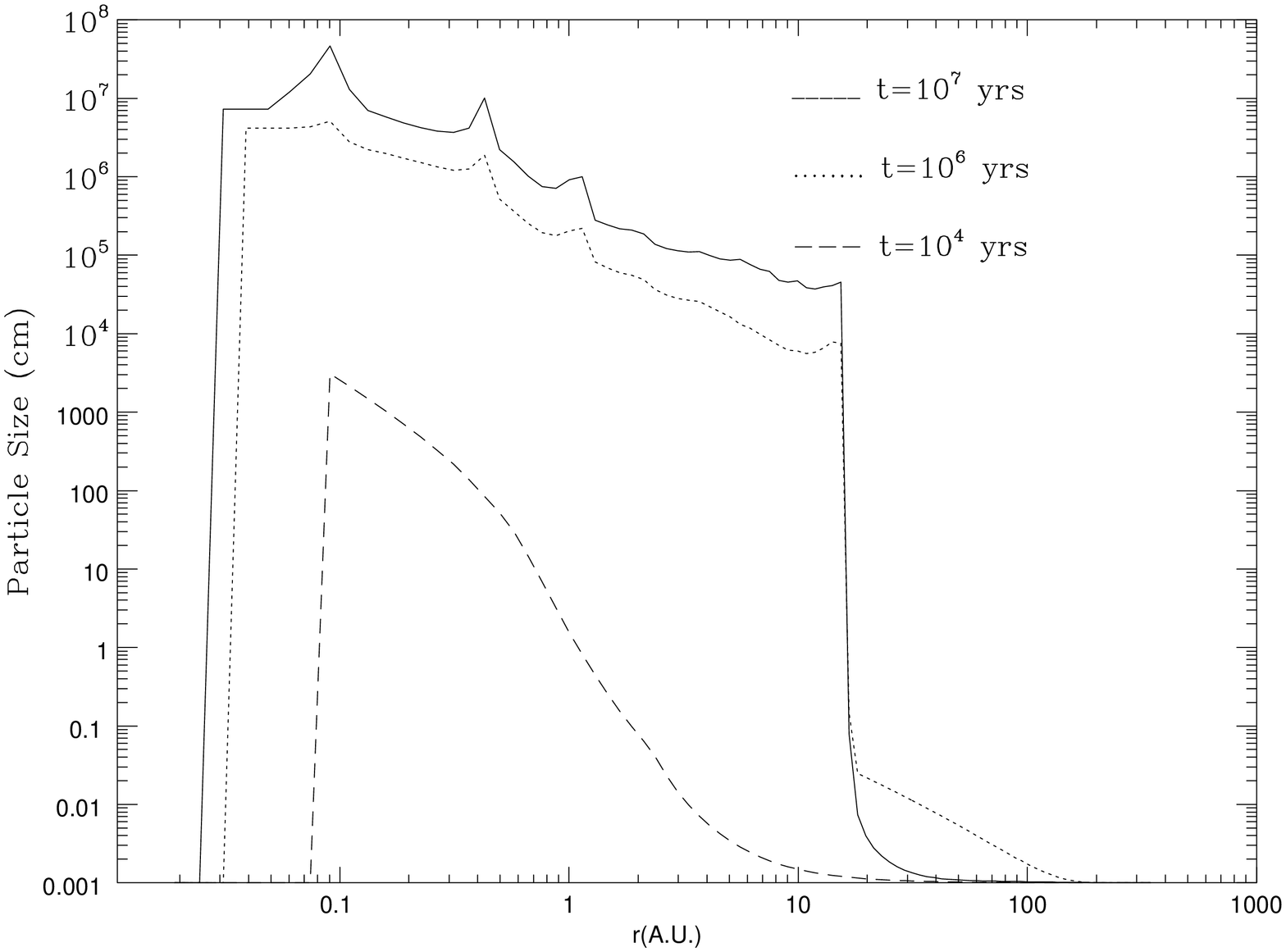,width=7cm,angle=270} (b)
\hspace{1cm}
\psfig{figure=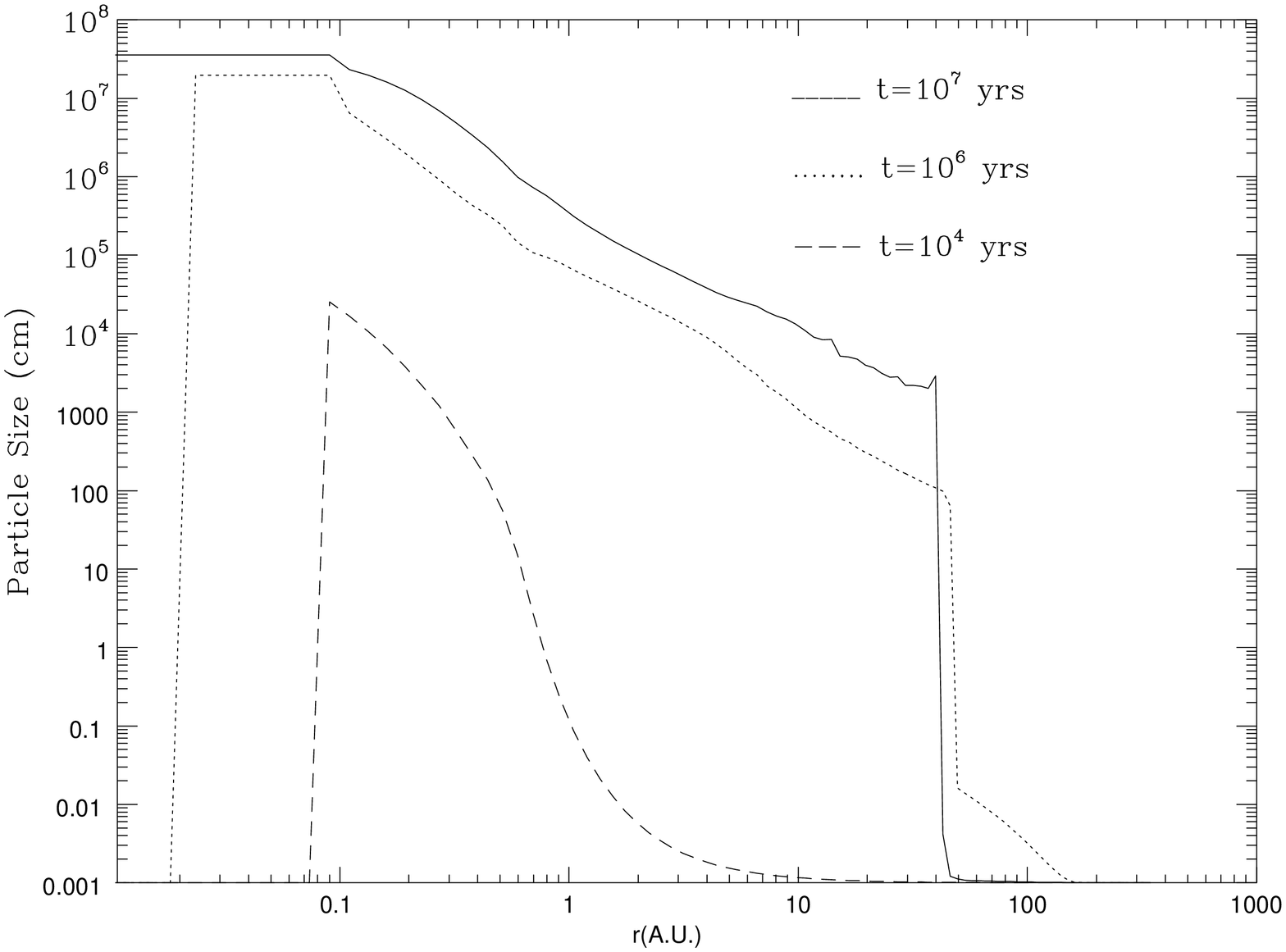,width=7cm,angle=270} 
}}
\caption[]{Evolution of the particle size in a disk of total gas mass
$M_{\rm d}=0.01 M_{\odot}$ with $\alpha=0.01$, panel (a), and
$\alpha=0.0001$, panel (b), at $t=10^4 {\rm yr}$ (dashed line),
$t=10^6 {\rm yr}$ (dotted line) and $t=10^7 {\rm yr}$ (solid line).}
\label{Figsize}
\end{figure*}

In this article, similarly to DP1 and DP2, we are mainly
interested in studying the planet migration due to the interaction
with planetesimals. For this reason we assume that the gas is almost
dissipated when the planet starts its migration.
\footnote{Clearly the effect of the presence of gas should be that of
accelerating the loss of angular momentum of the planet and to reduce
the migration time. In our case, there is still gas after $10^7 {\rm yrs}$, but it is in  
quantity inferior to that of planetesimals, expecially in the case of lower mass discs, in 
which it can be even two order of magnitudes less than planetesimals. Moreover, as noticed by 
Kominami \& Ida (2002) dynamical friction and gravitational gas drag are essentially the 
same dynamical process and then the effect of the gravitational gas drag is incorporated in 
that of dynamical friction. Other forces, as aerodynamical gas drag can be neglected when 
compared to gravitational gas drag (Kominami \& Ida 2002), in our case}  Moreover we know that the behavior of gas and
dust/planetesimals is different. Usually, the decline of gas mass near
stars is more rapid than the decline of the mass of orbiting solid
matter \citep{Zuckerman1}.  While the gas tends to be dissipated,
(several evidences show that the disk lifetimes range from $10^5$ yr
to $10^7$ yr, see \citealt{Strom1}; \citealt{Ruden1}), the coagulation
process induces an increase of the density of solid particles with
time (see Figs.\ref{FigM1solid}-\ref{FigM4solid}) and gives rise to
objects of increasing dimensions, as shown in Fig.\ref{Figsize}.

We clearly see that after times of the order of $10^6-10^7 {\rm yr}$,
the coagulation process gives rise to large particles ($>10^5 {\rm
cm}$) which have a small radial velocity, whence a negligible radial
motion.  This leads to a freezing of the solid surface density to the
value reached at times of the order of $10^7 {\rm yr}$.  In other
terms, once solids are in the form of planetesimals, the gas coupling
becomes unimportant and the radial distribution of solids does not
change any more. This is why we do not need to calculate its evolution
for times longer than $10^7$ years.
\footnote{Note however that the size distribution of planetesimals
keeps changing due to their mutual gravitational interaction.}  Then
the disk is populated by residual planetesimals for a longer period.
Here it is important to stress that planetesimal formation is not
independent from initial conditions. In particular, the final solid
surface density depends in an intricate fashion on the initial disk
mass $M_{\rm d}$ and on the turbulent viscosity parameter $\alpha$, see
\citet{SV2} and \citet{Kornet1}.

In order to investigate the dependence of the giant planet migration
on the properties of the protoplanetary disk we integrated the model
introduced in the previous sections for several values of the initial
disk surface density (i.e. several disk masses), and
different values of $\alpha$. More precisely, as in \citet{SV2} we
consider an initial gas surface density of the form:
\be
\Sigma_0(r)= \Sigma_1 \left[1+(r/r_1)^2\right]^{-3.78} + \Sigma_2
(r/1{\rm AU})^{-1.5} .
\label{Sigma0}
\ee
The quantities $\Sigma_1$, $r_1$ and $\Sigma_2$ are free parameters
which we vary in order to study different disk masses. The values we
use are given in Tab.1, where $M_{\rm d}$ is the gas disk mass (in units of
$M_{\odot}$), $J_{50}$ is the disk angular momentum (in units of
$10^{50}$ g cm$^2$s$^{-1}$), $\Sigma_1$ and $\Sigma_2$ are in g
cm$^{-2}$ and $r_1$ is in AU. The first term in Eq.(\ref{Sigma0})
ensures that there is some mass up to large distances from the star,
while the second term corresponds to the central concentration of the
mass and sets the location of the evaporation radius. Note however
that in any case the evaporation radius for the high-temperature
silicates we study here remains of order $0.1$ AU. As explained in
Sect.2.1 we consider only one species of solids: high-temperature
silicates with $T_{\rm evap}=1350$ K. We initialize the dust subdisk
at time $t=10^4$ yr (i.e. after the gas distribution has relaxed
towards a quasi-stationary state) by setting the solid surface density
$\Sigma_{\rm s}$ as: $\Sigma_{\rm s} = 6 \times 10^{-3} \Sigma$ in
order to account for cosmic abundance.

\begin{table} 
\begin{center} 
\caption{Properties of the initial gas disk}
\begin{tabular}{|l|l|l|l|l|} \hline
$M_{\rm d}$      &  $J_{50}$  &  $\Sigma_1$ &  $r_1$  &  $\Sigma_2$ \\ \hline
$10^{-1}$  &  $911$     &  $22$       &  $200$  &  $100$      \\ \hline
$10^{-2}$  &  $85$      &  $1.7$      &  $200$  &  $100$      \\ \hline
$10^{-3}$  &  $5.5$     &  $1.2$      &  $50$   &  $30$       \\ \hline
$10^{-4}$  &  $0.46$    &  $0.2$      &  $50$   &  $2.8$      \\ \hline
\end{tabular}
\end{center} 
\label{table1} 
\end{table}

\subsection{Evolution of the gas disk}

\begin{figure*}
\centerline{\hbox{(a)
\psfig{figure=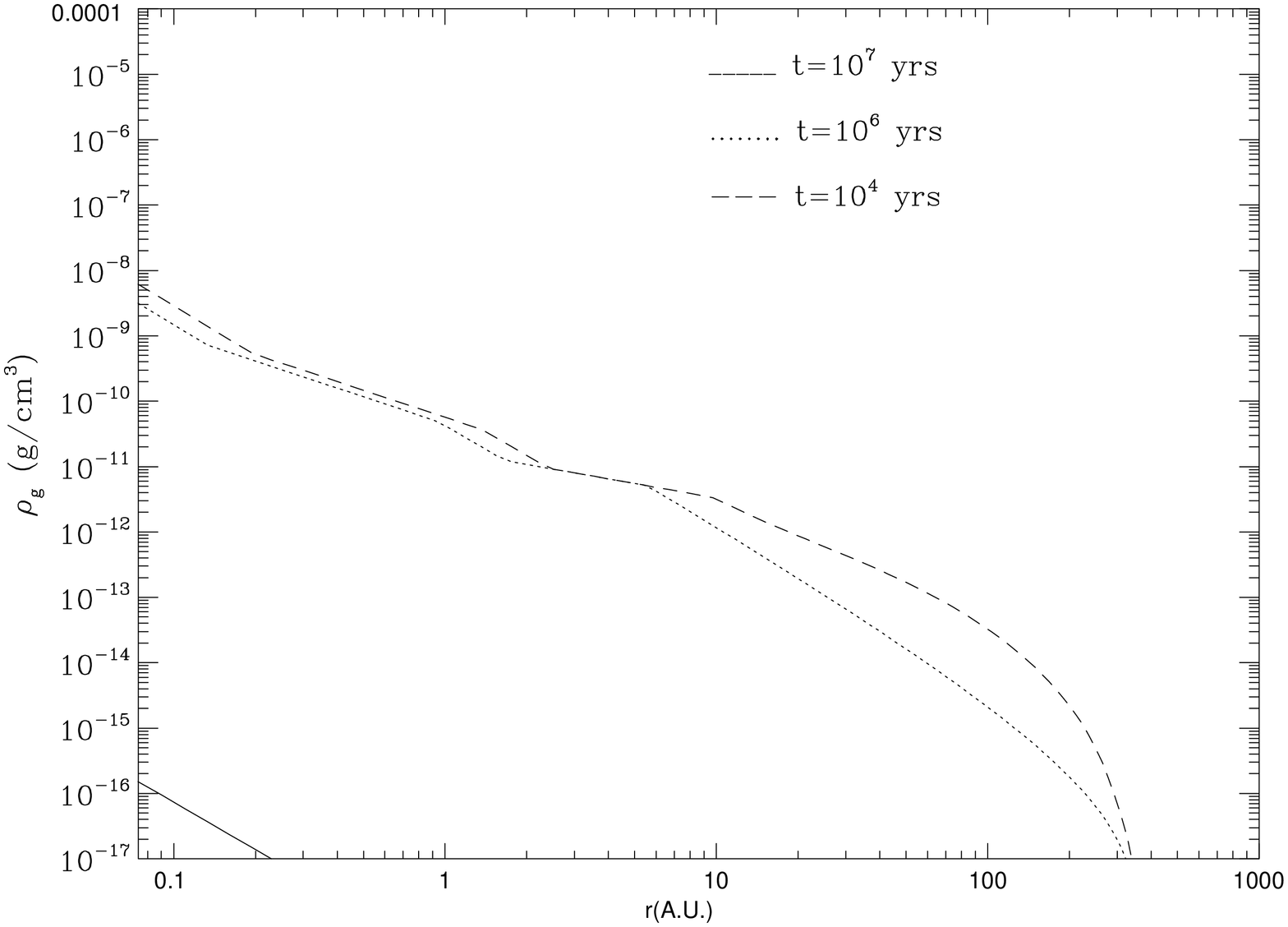,width=6cm,angle=270} (b)
\hspace{1cm}
\psfig{figure=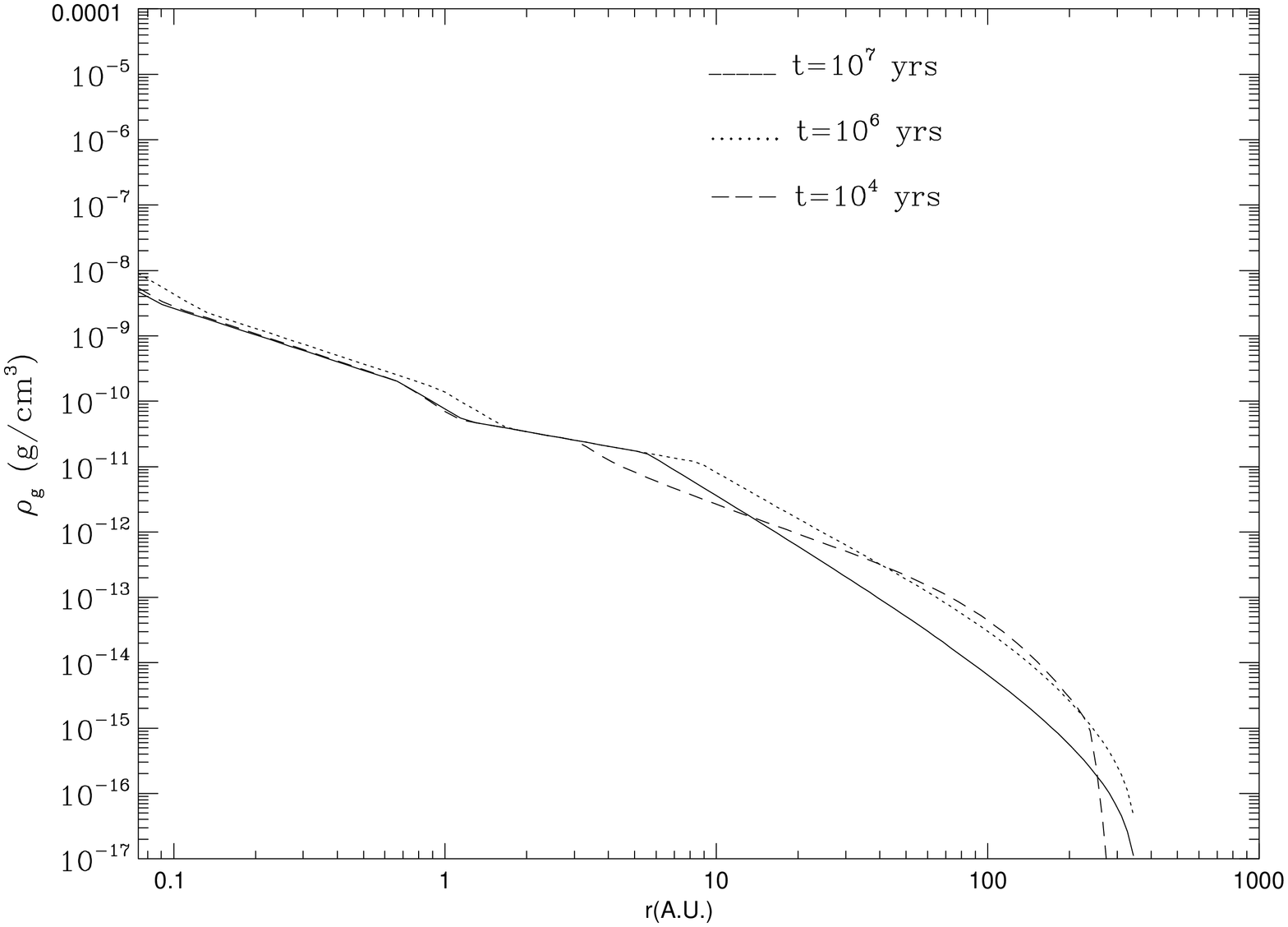,width=6cm,angle=270}
}}
\centerline{\hbox{(c)
\psfig{figure=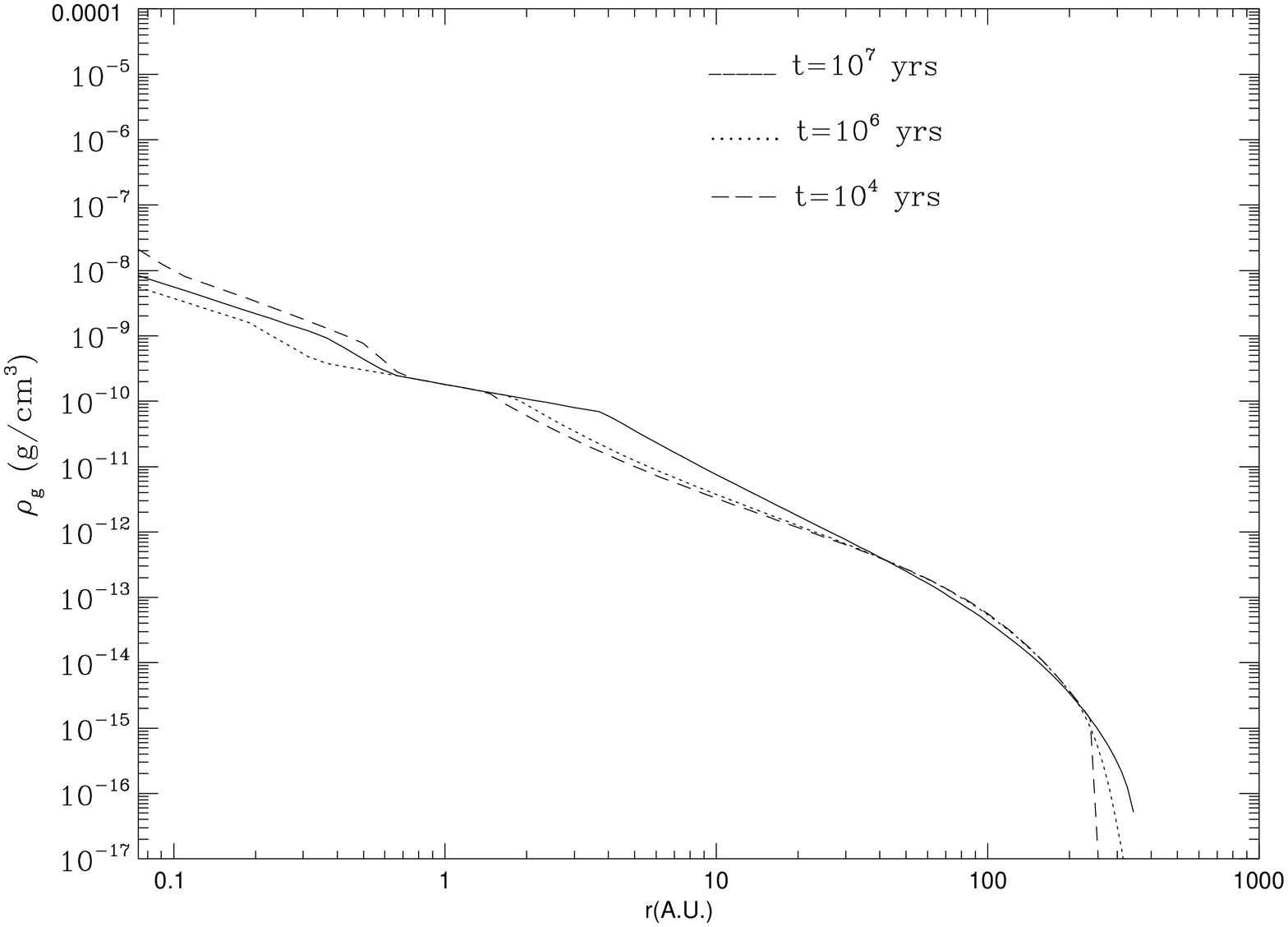,width=6cm,angle=270} (d)
\hspace{1cm}
\psfig{figure=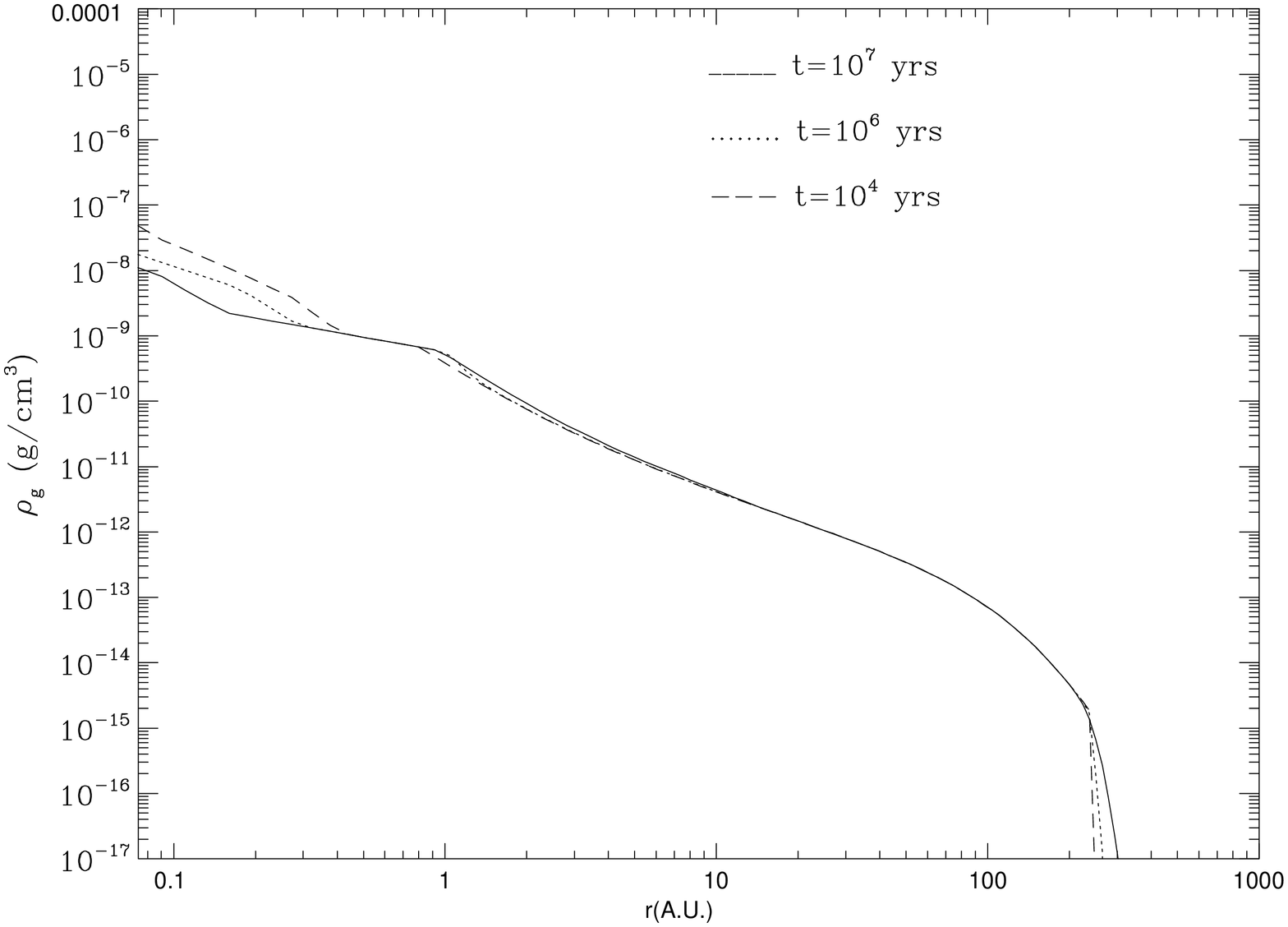,width=6cm,angle=270}
}}
\caption[]{(a) Evolution of gas midplane density $\rho$, for a disk
with $M_{\rm d}=0.1 M_{\odot}$ and $\alpha=0.1$ at $t=10^4 {\rm yr}$
(dashed line), $t=10^6 {\rm yr}$ (dotted line) and $t=10^7 {\rm yr}$
(solid line). (b) Same as Fig.2a but with $\alpha=0.01$. (c) Same as
Fig.2a but with $\alpha=0.001$. (d) Same as Fig.2a but with
$\alpha=0.0001$}
\label{FigM1gas}
\end{figure*}

\begin{figure*}
\centerline{\hbox{(a)
\psfig{figure=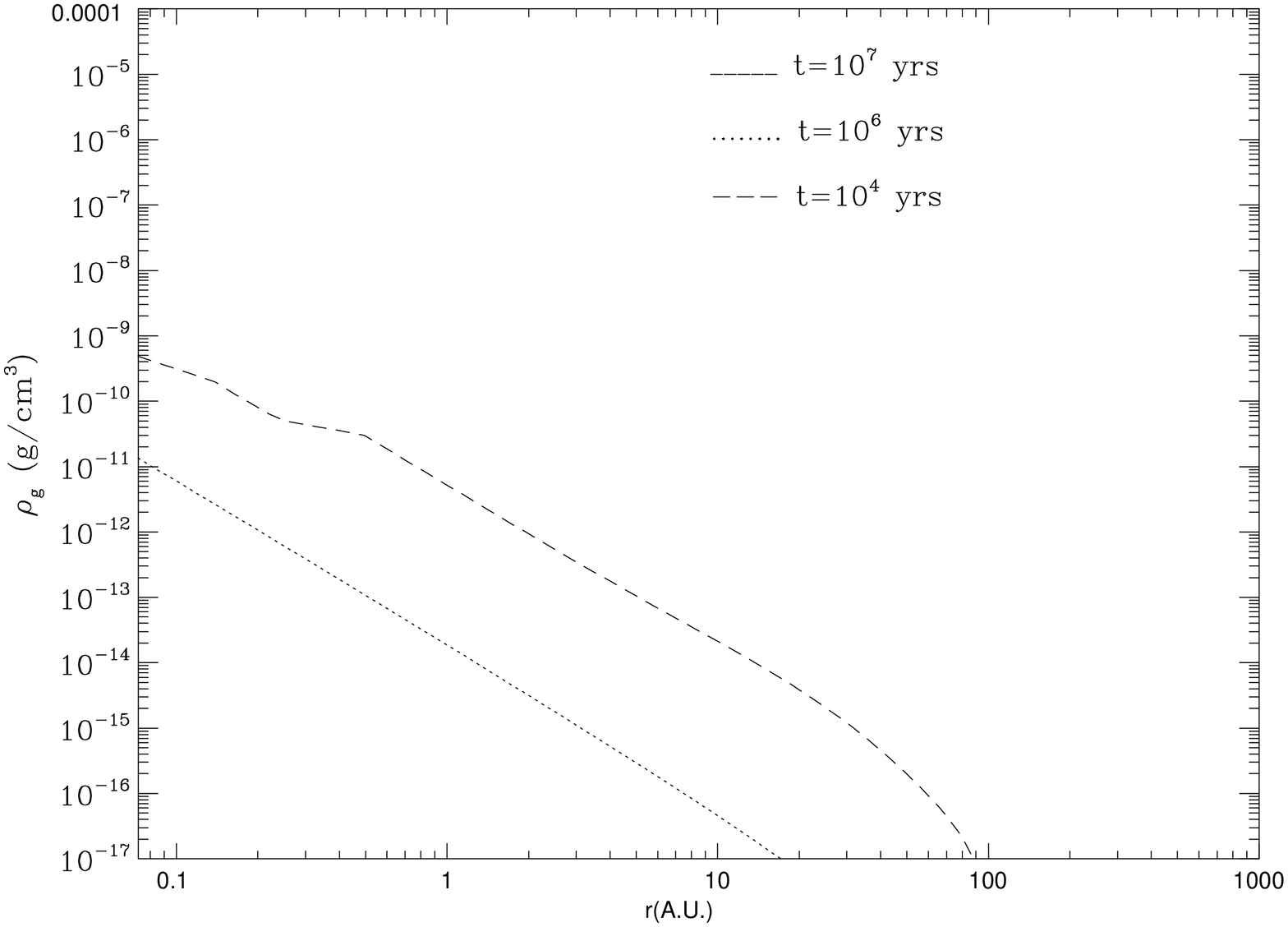,width=6cm,angle=270} (b)
\hspace{1cm}
\psfig{figure=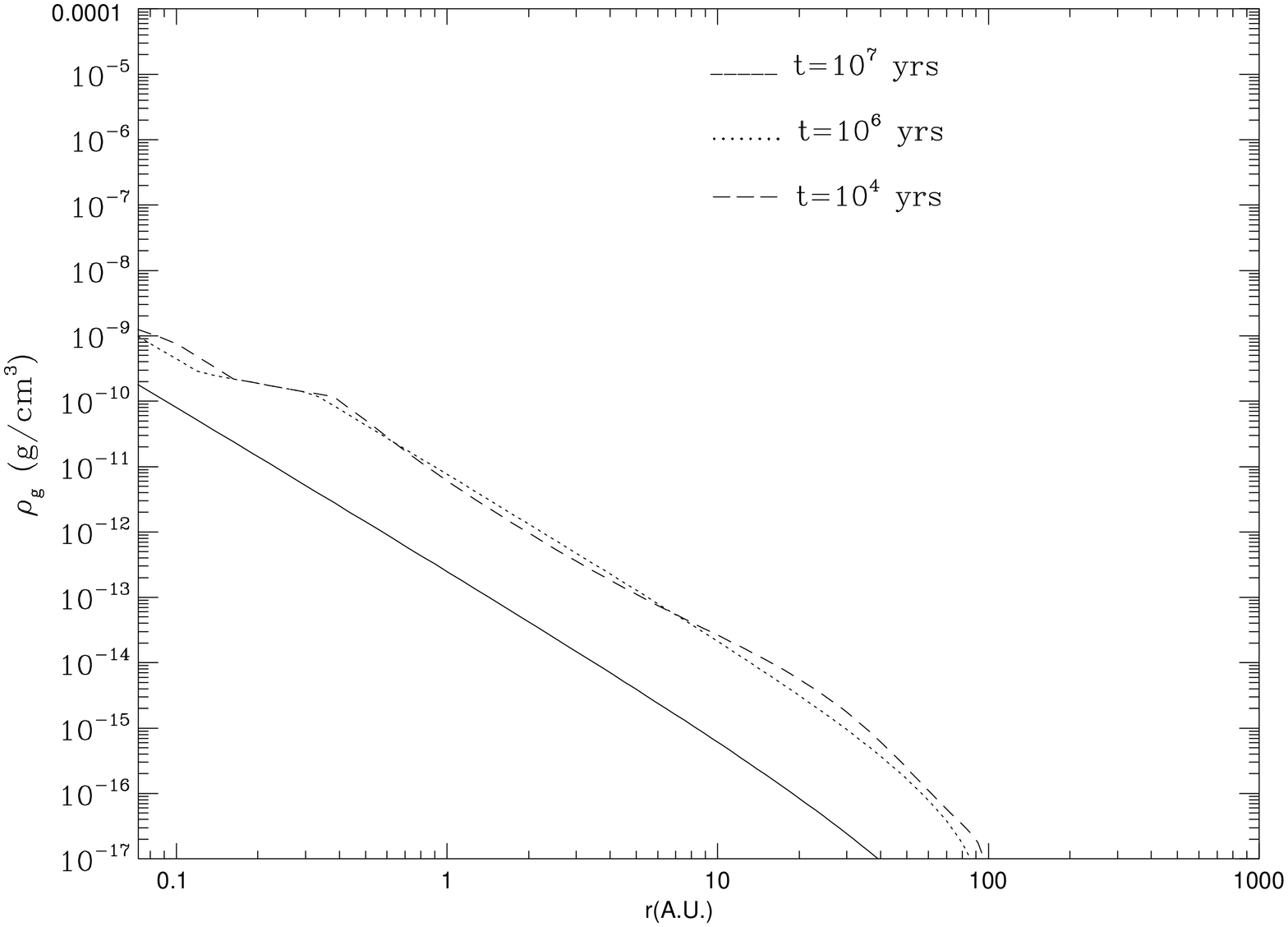,width=6cm,angle=270}
}}
\centerline{\hbox{(c)
\psfig{figure=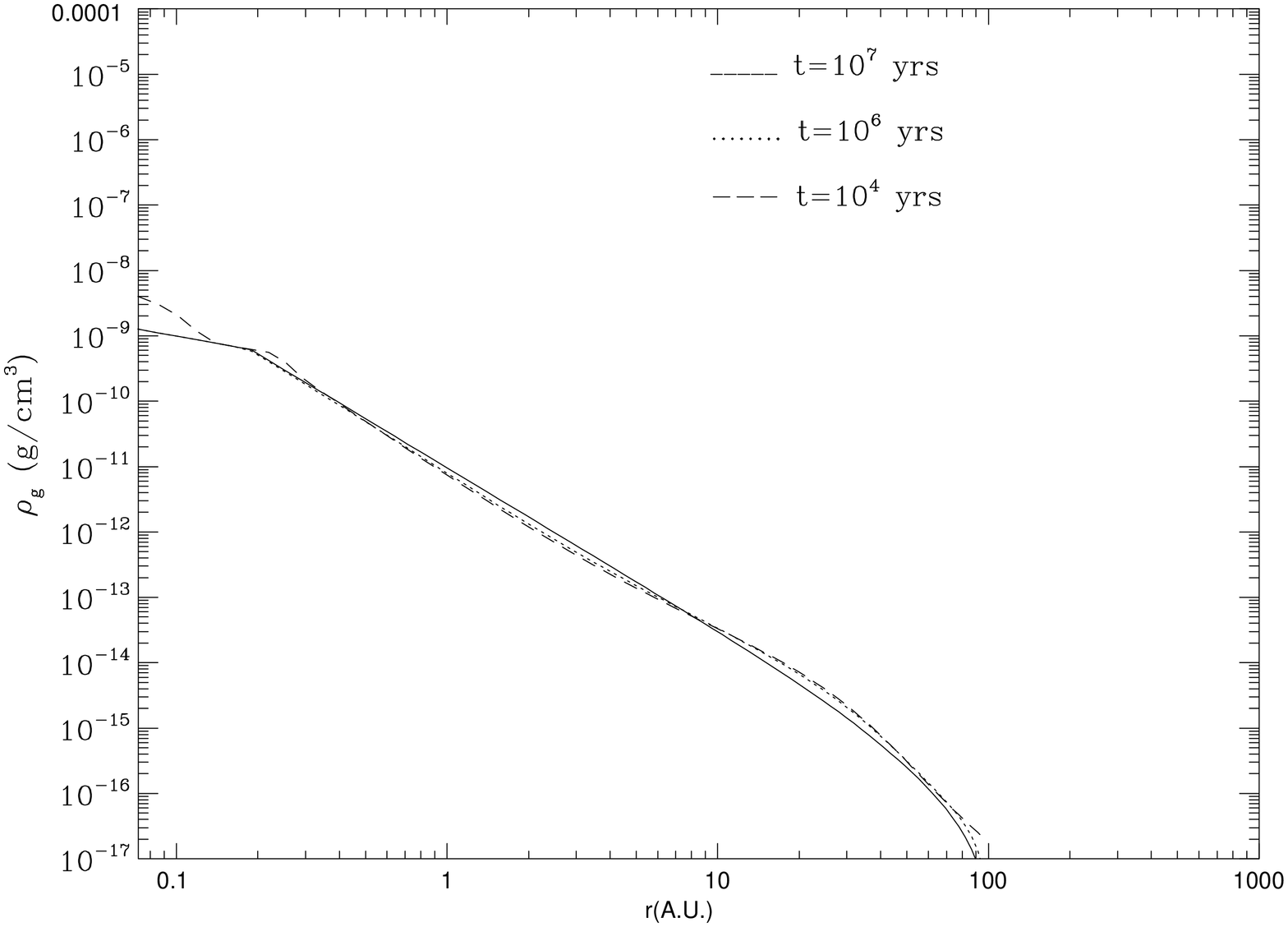,width=6cm,angle=270} (d)
\hspace{1cm}
\psfig{figure=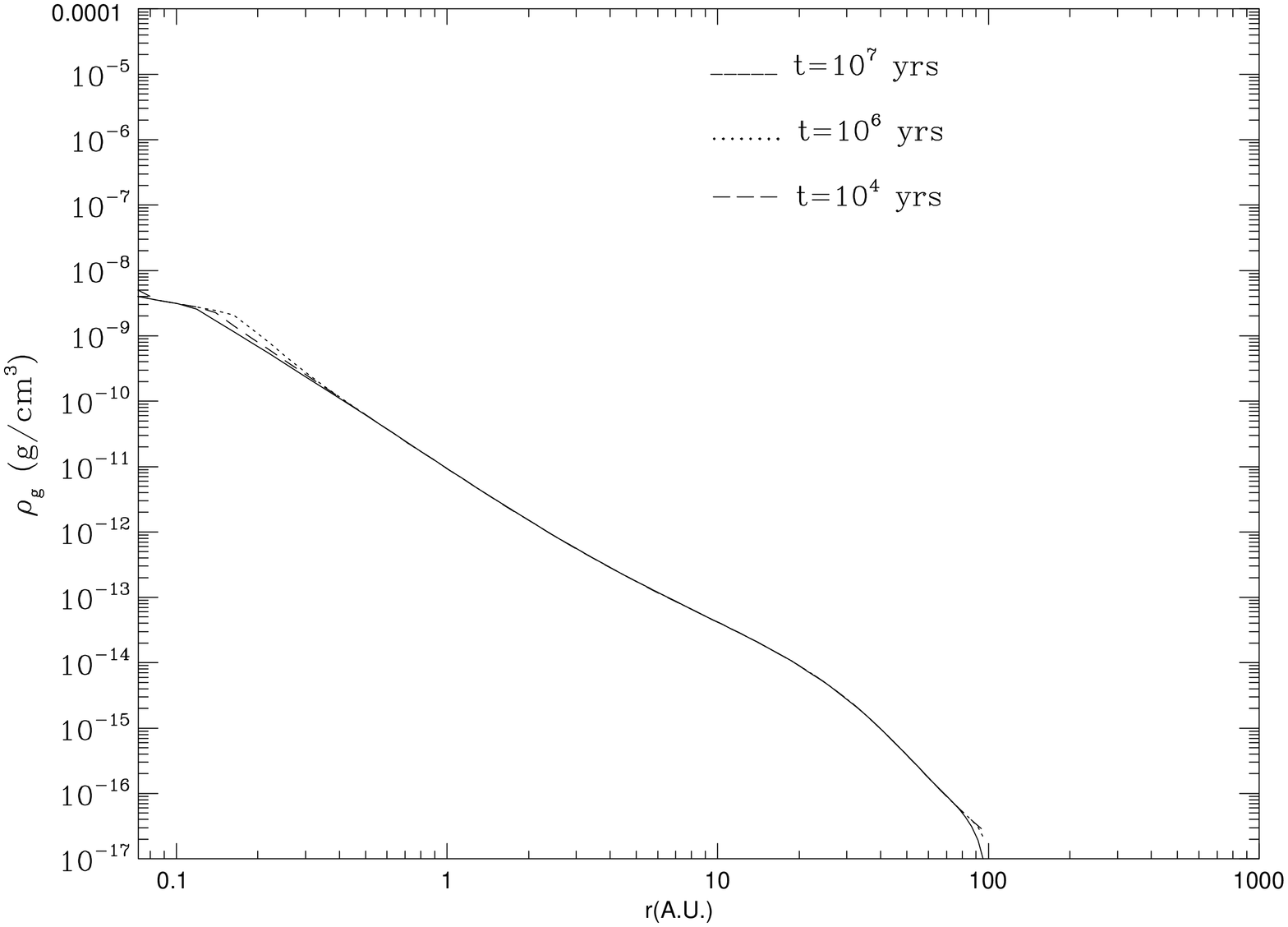,width=6cm,angle=270}
}}
\caption[]{(a) Evolution of gas midplane density $\rho$, for a disk
with $M_{\rm d}=0.0001 M_{\odot}$ and $\alpha=0.1$ at $t=10^4 {\rm
yr}$ (dashed line), $t=10^6 {\rm yr}$ (dotted line) and $t=10^7 {\rm
yr}$ (solid line). (b) Same as Fig.3a but with $\alpha=0.01$. (c) Same
as Fig.3a but with $\alpha=0.001$. (d) Same as Fig.3a but with
$\alpha=0.0001$}
\label{FigM4gas}
\end{figure*}

We show in Fig.\ref{FigM1gas} the evolution of the midplane gas
density $\rho$ for the disk of initial mass
$M_{\rm d}=0.1 M_{\odot}$, with four values of $\alpha$ from $10^{-1}$
down to $10^{-4}$. This range covers all values of $\alpha$ that are
conceivable in the context of protoplanetary disks. Of course, we can
check that the gas disk is rapidly depleted when the viscosity is
high. We show in Fig.\ref{FigM4gas} the results we obtain for the gas
in the case of a small mass disk: $M_{\rm d}=10^{-4}
M_{\odot}$. Obviously we recover the same overall trends. Note that
the main difference between our low-mass and high-mass scenarios shows
in the outer radius of the gas disk. Indeed, with the initial
conditions defined in Tab.1 massive disks are more extended (see the
characteristic radius $r_1$). On the other hand, the gas density below
$0.5$ AU does not change by much so that the evaporation radius is
always of order $0.1$ AU.

It is to note that in Figs. 3b,c,d, and Fig. 4c the solid line that 
corresponds to the longest time is, at places, above the other lines. Since the gas 
evolution is governed by a diffusion equation, one should expect that the  
surface density decreases with time at any point in the disk. 
%An important point to stress is the fact that 
In reality it is not always true that the surface density 
must decrease with time at 
any point in the disk, although its evolution is indeed given by a diffusion 
equation.
%
%%First, the disk should extend a bit outward, because of this diffusion and of
%%the outward transport of angular momentum. This is seen in our Fig. 3b and also
%%in Fig.4 in my paperI with Stepinski.
%
%Second, 
In fact, in the inner parts of the disk there is also a net mass flux inward
(so that the gas disk looses some mass onto the star). Then, depending on
the initial conditions and $\alpha$, if there is a lot of mass at large radii when
this matter diffuses down to a smaller radius it can lead to a local increase
of the surface density. This is what happens in Fig. 3b. 
%and also probably
%in Fig.5a in my paperII with Stepinski.
%Our Fig. 3b seems rather clear in this respect. 
Most of the mass initially lies at ~20-200 AU, then, this mass slowly diffuses 
both outward and inward. Thus,
one can check that in this range of radii the density indeed decreases with 
time. 
However, at very large radii the density increases somewhat (more 
precisely the outer radius of the disk grows). On the other hand, at small 
radii the density may also increase locally (but this is not always the case)
as a large amount of matter may flow inward.
Finally, in the quoted figures, we plot the density which makes the discussion a bit more
difficult since it depends on both $\Sigma$ and H (which decreases somewhat for
smaller $\Sigma$). However this should not change the basics of the arguments.

\subsection{Evolution of the solid subdisk}

\begin{figure*}
\centerline{\hbox{(a)
\psfig{figure=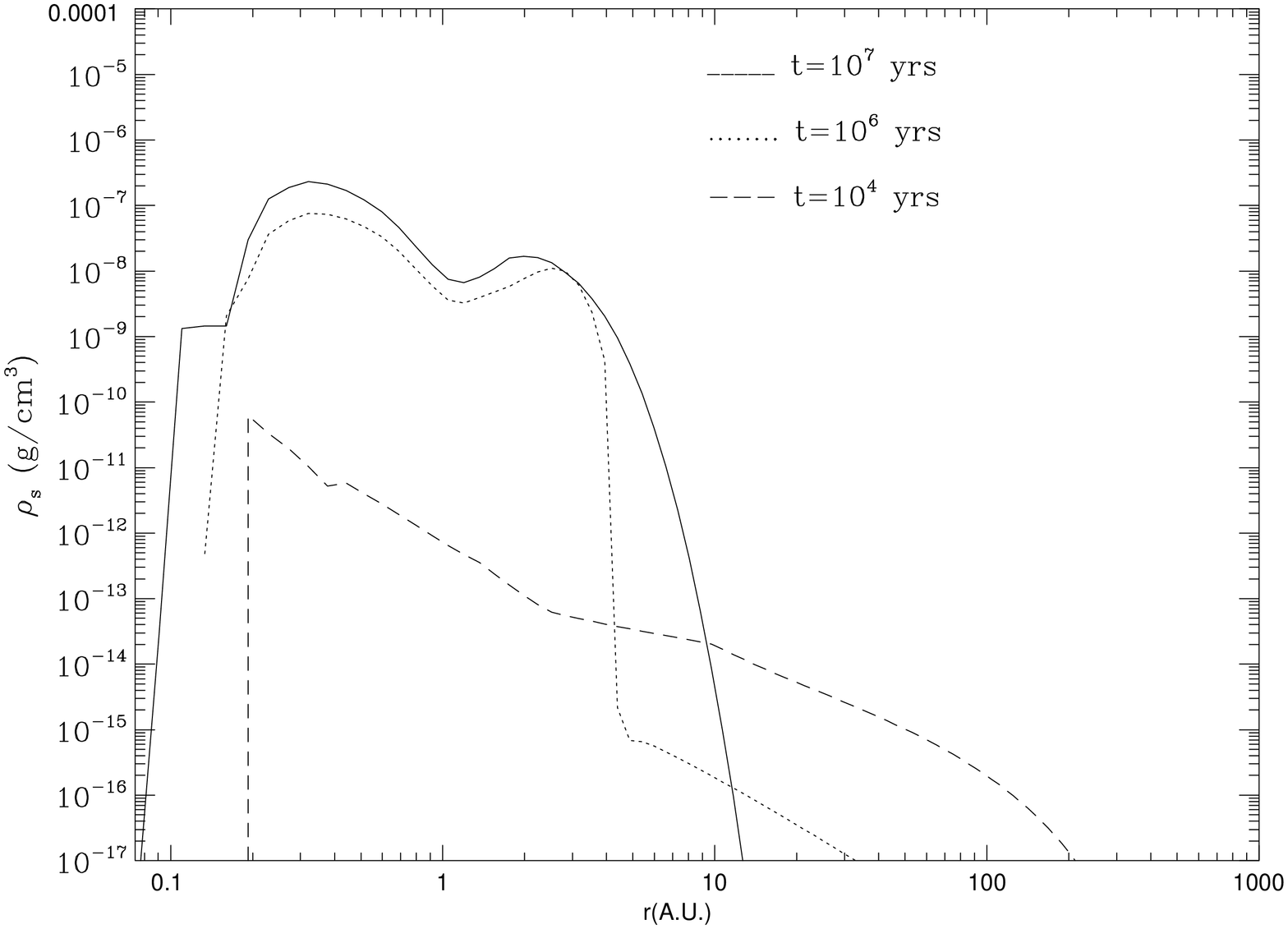,width=6cm,angle=270} (b)
\hspace{1cm}
\psfig{figure=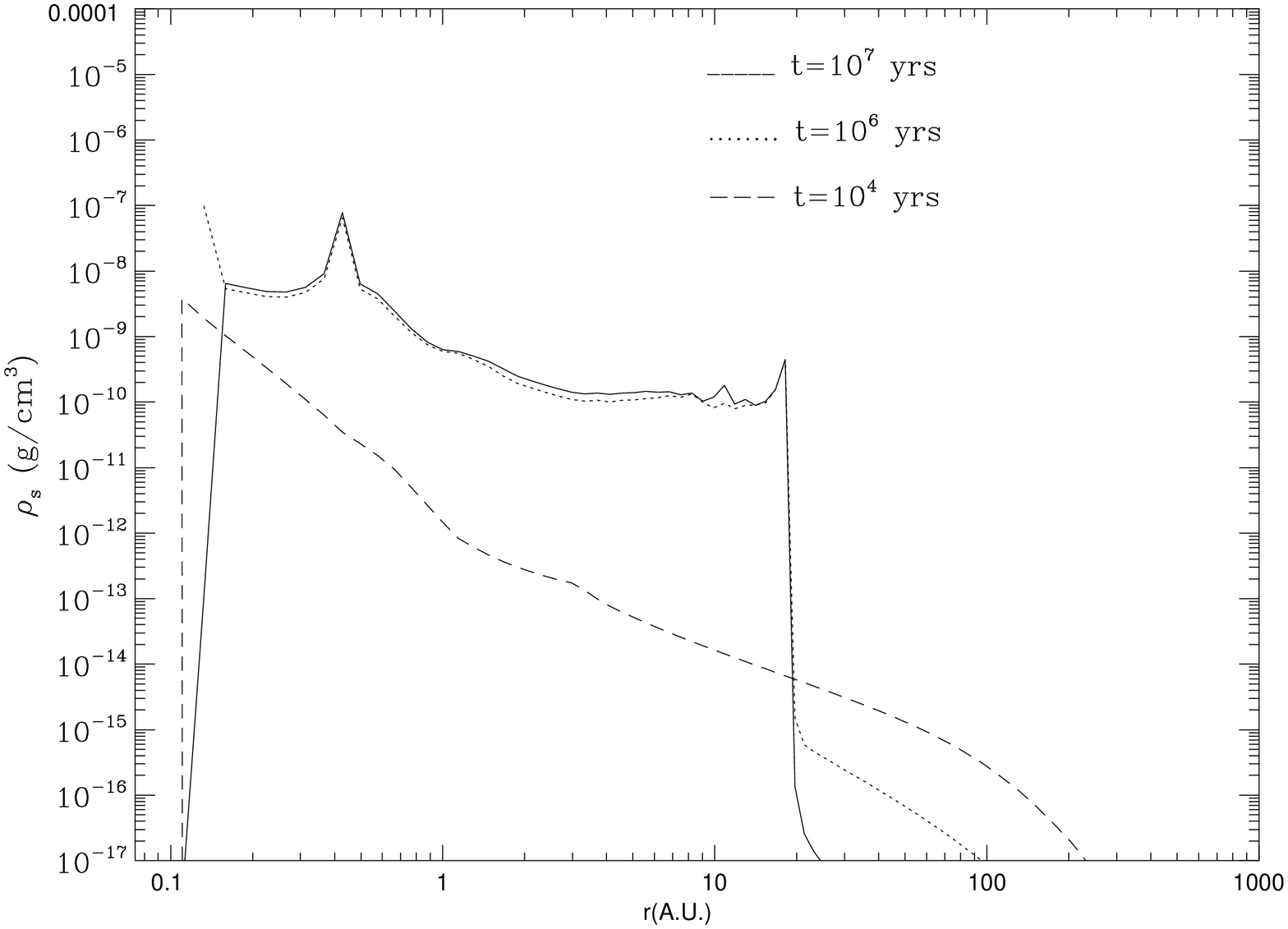,width=6cm,angle=270}
}}
\centerline{\hbox{(c)
\psfig{figure=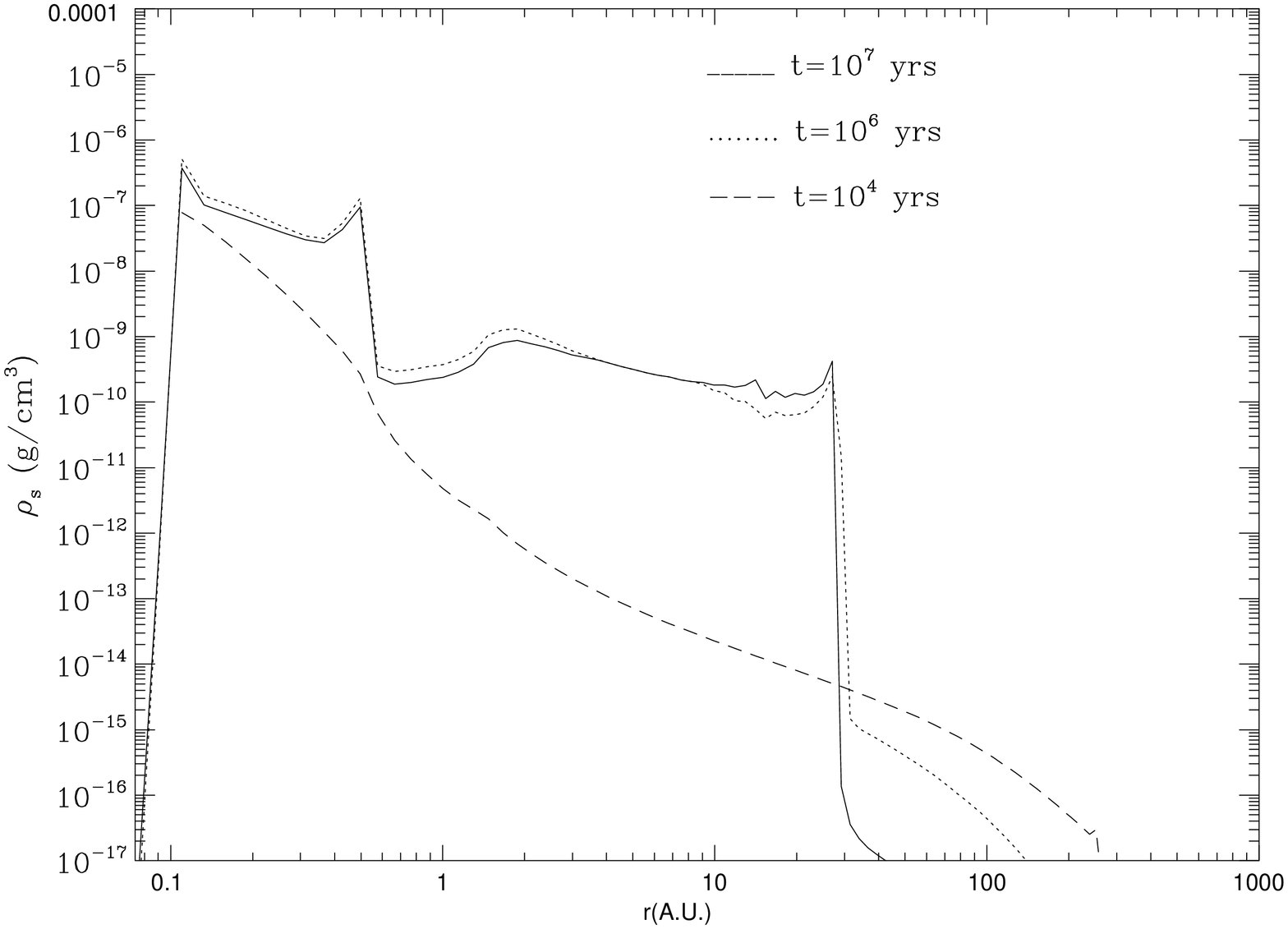,width=6cm,angle=270} (d)
\hspace{1cm}
\psfig{figure=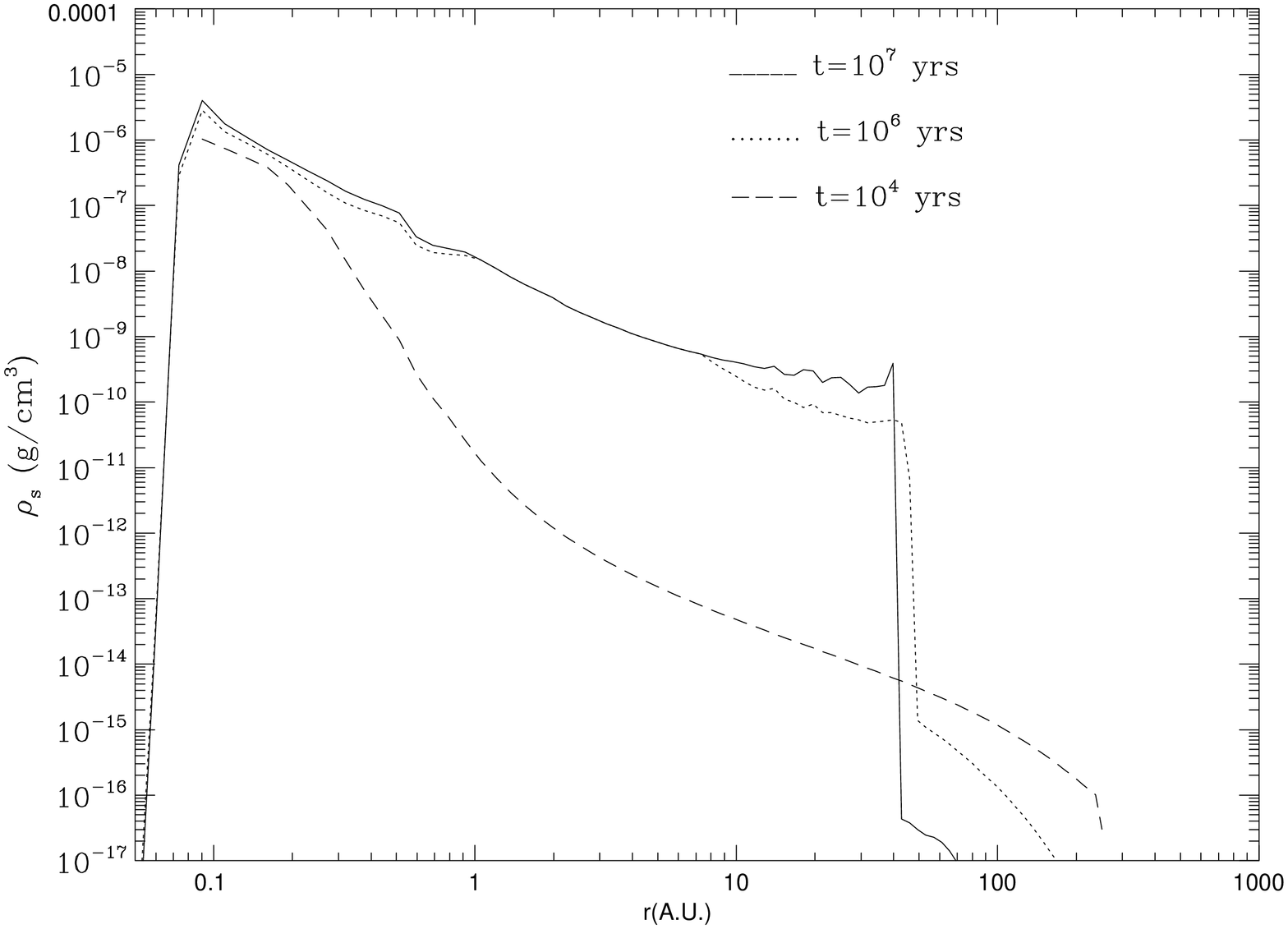,width=6cm,angle=270}
}}
\caption[]{(a) Evolution of solids midplane density $\rho_{\rm s}$ for
a disk with $M_{\rm d}=0.1 M_{\odot}$ and $\alpha=0.1$ at $t=10^4 {\rm
yr}$ (dashed line), $t=10^6 {\rm yr}$ (dotted line) and $t=10^7 {\rm
yr}$ (solid line). (b) Same as Fig.4a but with $\alpha=0.01$. (c) Same
as Fig.4a but with $\alpha=0.001$. (d) Same as Fig.4a but with
$\alpha=0.0001$}
\label{FigM1solid}
\end{figure*}

\begin{figure*}
\centerline{\hbox{(a)
\psfig{figure=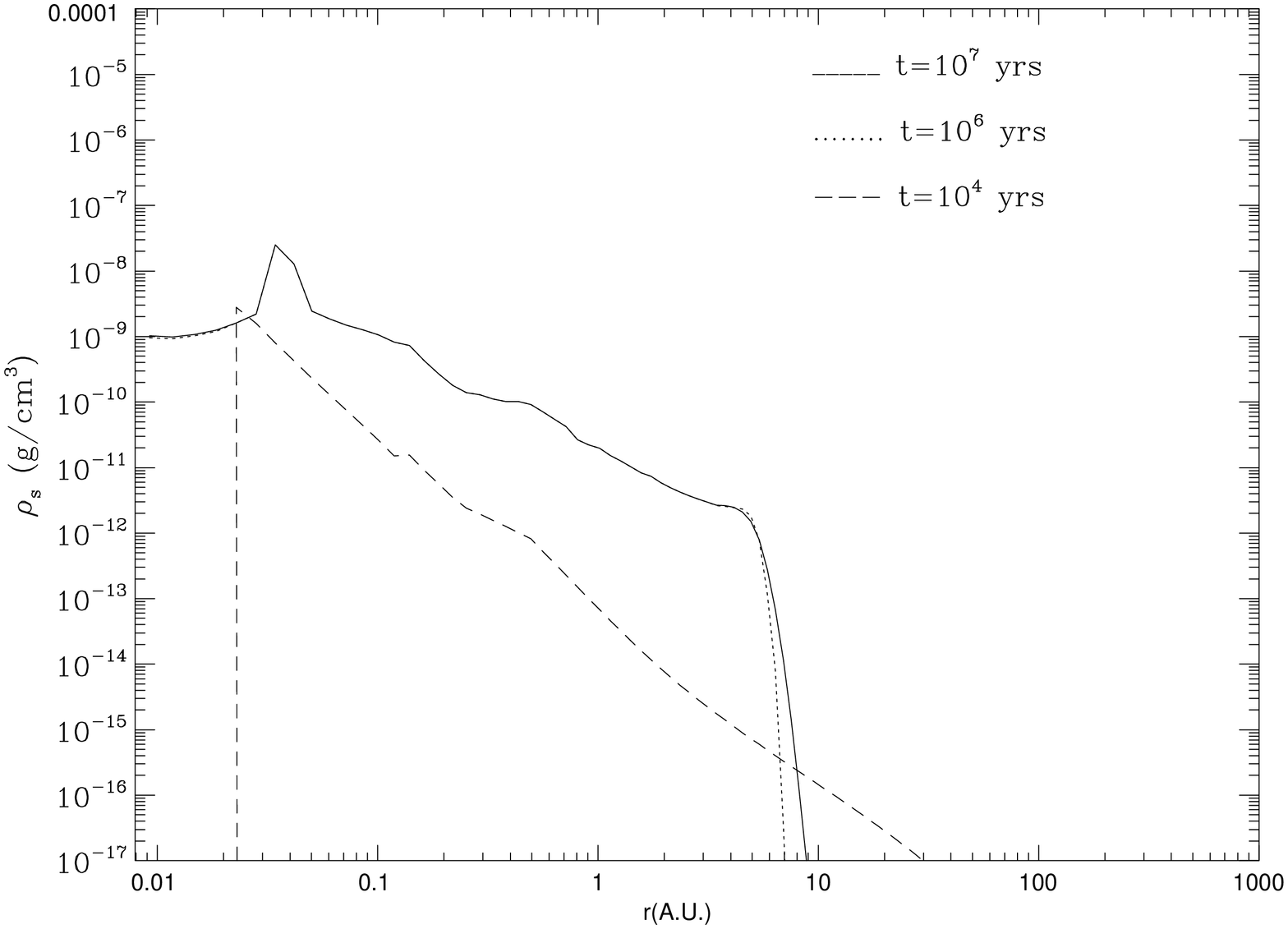,width=6cm,angle=270} (b)
\hspace{1cm}
\psfig{figure=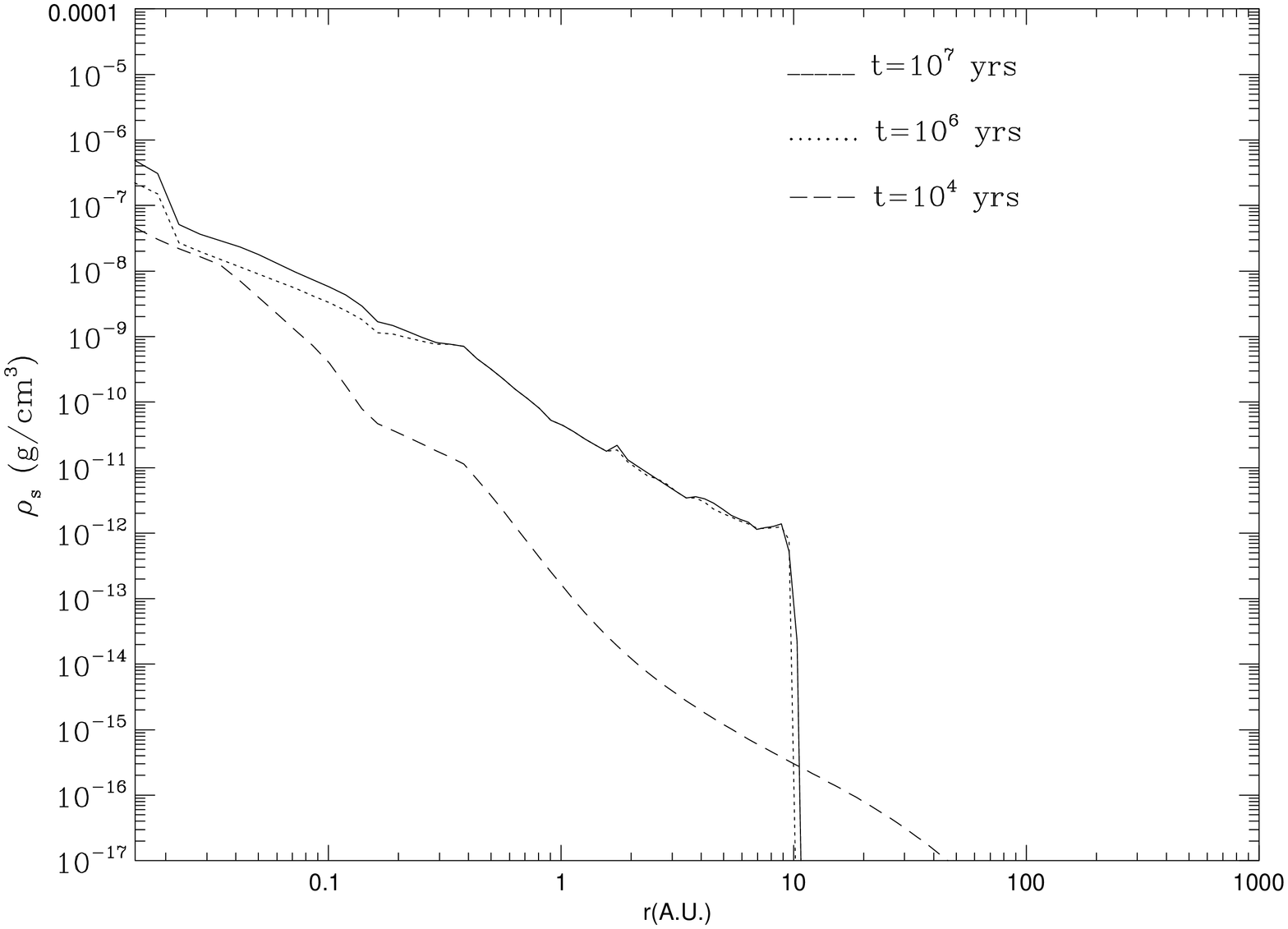,width=6cm,angle=270}
}}
\centerline{\hbox{(c)
\psfig{figure=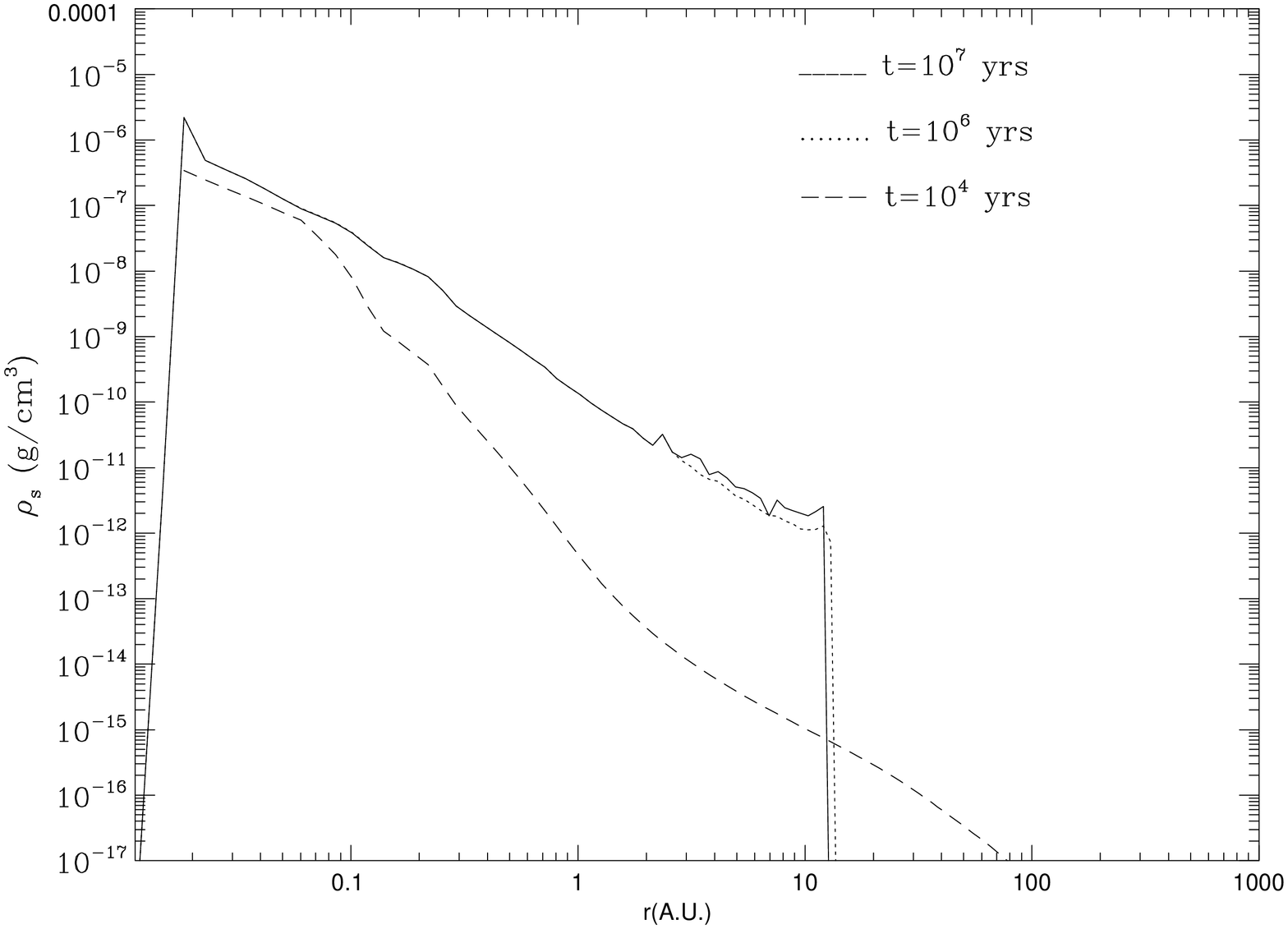,width=6cm,angle=270} (d)
\hspace{1cm}
\psfig{figure=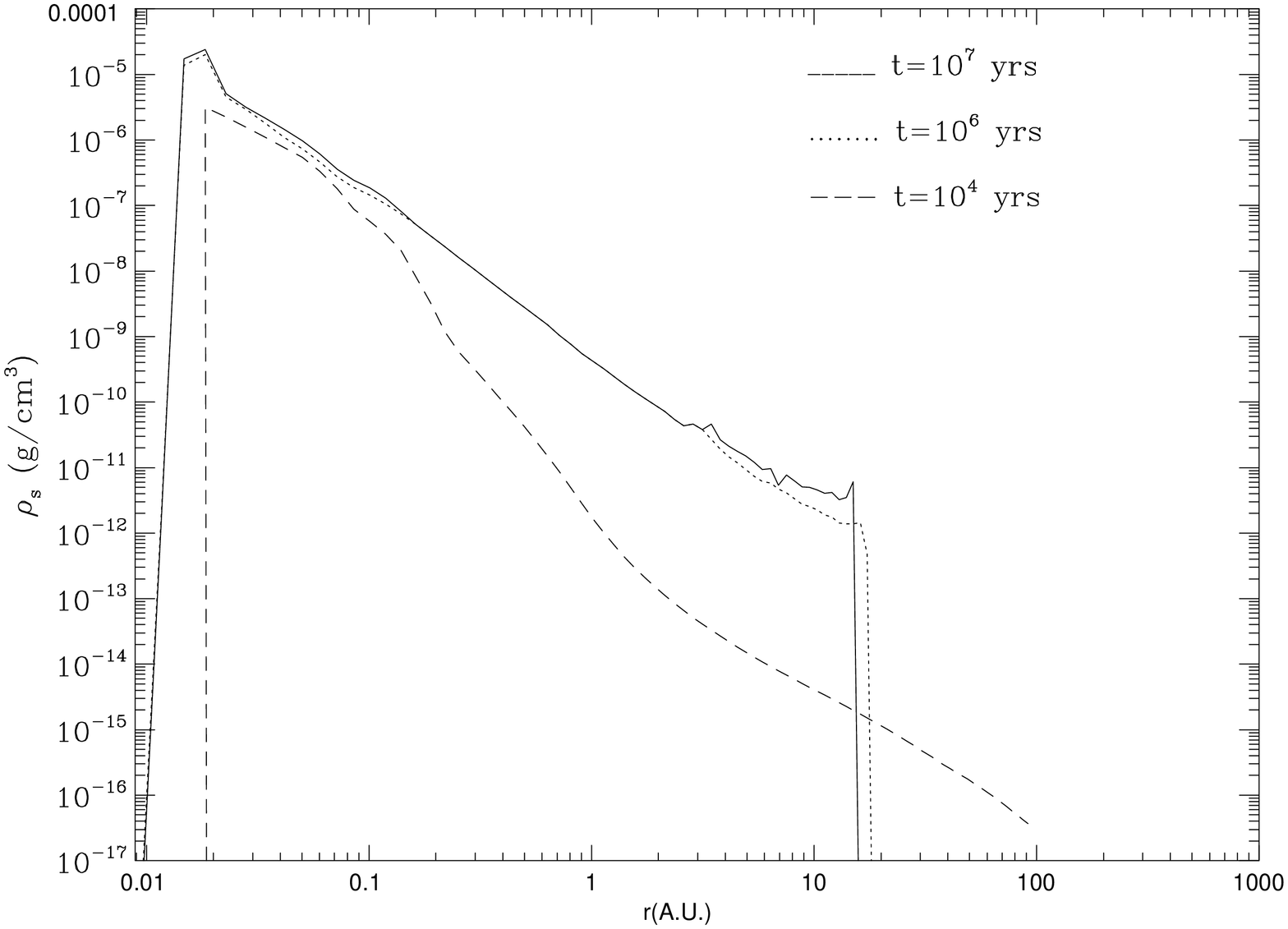,width=6cm,angle=270}
}}
\caption[]{(a) Evolution of solids midplane density $\rho_{\rm s}$ for
a disc with $M_d=0.0001 M_{\odot}$ and $\alpha=0.1$ at $t=10^4 {\rm
yr}$ (dashed line), $t=10^6 {\rm yr}$ (dotted line) and $t=10^7 {\rm
yr}$ (solid line).  (b) Same as Fig.5a but with $\alpha=0.01$.  (c)
Same as Fig.5a but with $\alpha=0.001$.  (d) Same as Fig.5a but with
$\alpha=0.0001$}
\label{FigM4solid}
\end{figure*}

Next, we show in Fig.\ref{FigM1solid} and Fig.\ref{FigM4solid} the
evolution of the solid midplane density for the two cases $M_{\rm d}=10^{-1}
M_{\odot}$ and $M_{\rm d}=10^{-4} M_{\odot}$, and four values of
$\alpha$. We can see a converged radial distribution of solids emerge
at late times of order $10^6$ yr. Note that although the radial
distribution of planetesimals depends on the value of $\alpha$, the
total mass of solids in a disk is roughly independent of $\alpha$
since it remains approximately equal to the initial mass of solids in
the disk. This means that solids are reshuffled within the disk but
they are not lost into the star.
\footnote{In fact, solids initially located close to the evaporation
radius are lost but they constitute a small percentage of the total
solid material, which is predominantly located in the outer disk.}
The value of $\alpha$ determines the radial distribution of solids:
particles in a disk with a larger value of $\alpha$ (more turbulent
disk) have larger inward radial velocities and consequently are locked
into planetesimals closer to the star than particles in a less
turbulent disk. Thus, the smaller the value of $\alpha$ the broader
the final distribution of solids.  The evaporation radius is located
at $\simeq 0.1 AU$ while the outer limit of converged $\rho_{\rm s}$
is about $10$ AU for $\alpha=10^{-1}$ and moves to $\simeq 70$ AU for
$\alpha=10^{-4}$, in the case $M_{\rm d}=0.1 M_{\odot}$. 
In the case $M_{\rm d}=0.0001
M_{\odot}$, the outer radius moves only from $\simeq 10$ to $\simeq
20$ AU, since the initial extension of the disk is smaller.

The coagulation process gives rise to solids of $10^6-10^7$ cm.  In
disks characterized by smaller values of $\alpha$, and thus a more
extended distribution of solids, the range of sizes goes from
$10^6-10^7$ cm at the evaporation radius down to $10^3-10^4$ cm at the
outer limit (see Fig.\ref{Figsize}). This is because the coagulation
process is less efficient at larger radii where the solid density is
smaller and the velocity dispersion of the dust decreases (along with
the gas temperature which governs the turbulent velocity). At later
times these solids will continue to increase their sizes, but will not
change their radial position, as they are already large enough to have
a negligible radial motion.  Note that the converged radial
distribution of solids does not vary monotonically. There are bulges
of matter near the evaporation radius as well as at the outer limit of
$\rho_{\rm s}$ (or $\Sigma_{\rm s}$). These bulges are present in all
cases, but become narrower for smaller values of $\alpha$. Their
existence is a consequence of an intricate and nonlinear
interdependence between advection and coagulation, modulated by
changing gas conditions and the character of turbulence. The formation
of the inner bulge also signals the ``freezing'' of the total mass of
the solids in the disk, inasmuch as no more solids are lost to the
vapor zone in subsequent disk evolution. They will either be captured
by the bulge or come to rest by themselves at larger radii. Notice
that it is a radial squeezing due to particle dynamics, rather than
vertical squeezing due to sedimentation, that is primarily responsible
for establishing the bulge and keeping particles from falling into the
vapor zone. The same mechanism is responsible for the abrupt drop in
$\rho_{\rm s}$, which we associate with the outer limit of solid
matter distribution, as well as the presence of the $\rho_{\rm s}$
bulge at the location of this drop.

Summarizing, the results show that the distribution of matter in the
disk is sensitive to the assumed initial conditions. After times
shorter than $3.2 \times 10^6$ yr the radial distribution of
planetesimal mass density $\rho_{\rm s}$ settles and can only change
on a much longer timescale by processes not considered in our model
(like mutual gravitational interactions between planetesimals proposed
first by \citealt{Safronov1}).
\footnote{Note that $10^6$ yr is the time required for $\Sigma_{\rm
s}$ (or $\rho_{\rm s}$) to converge everywhere. However, this
convergence is not uniform and can be achieved on a time scale as
short as $10^4$ yr in the innermost disk where the dust density is
highest.}  We refer the reader to \citet{SV2} and \citet{Kornet1} for
more detailed discussions of the behavior of protoplanetary disks.

\subsection{Migration of a giant planet}

\begin{figure*}
\centerline{\hbox{(a)
\psfig{figure=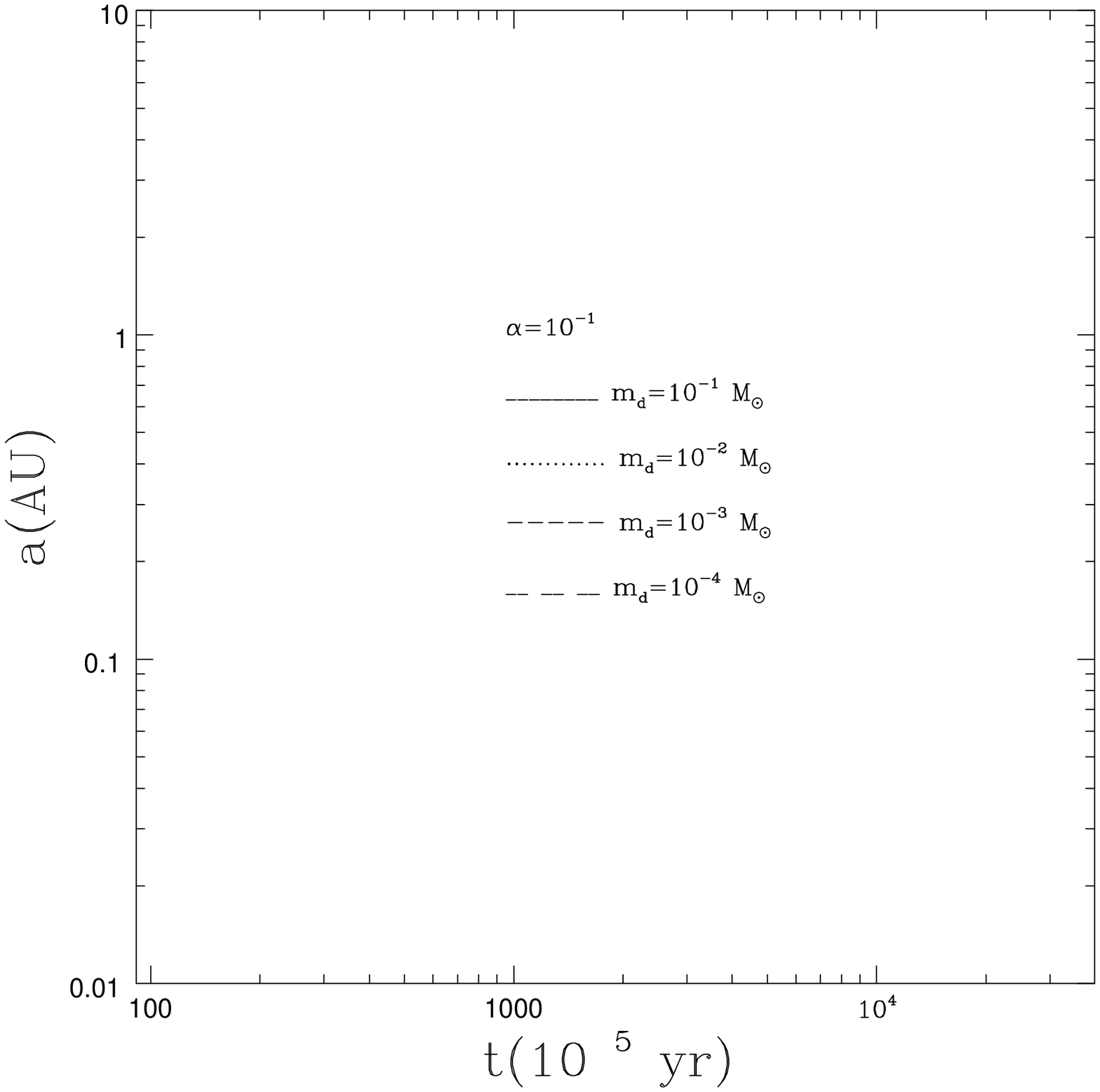,width=8cm} (b)
\vspace{1cm}
\psfig{figure=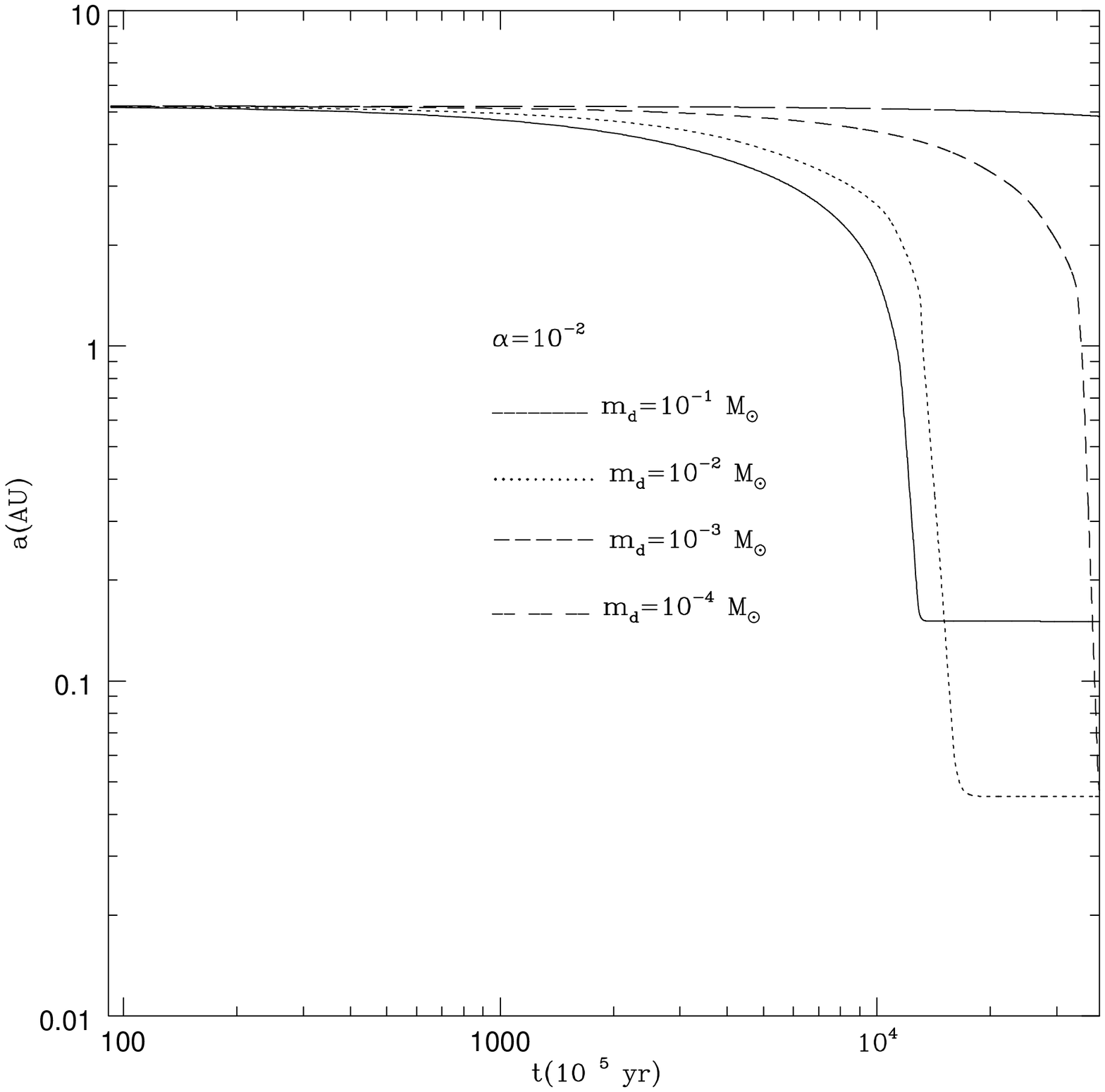,width=8cm}
}}
\centerline{\hbox{(c)
\psfig{figure=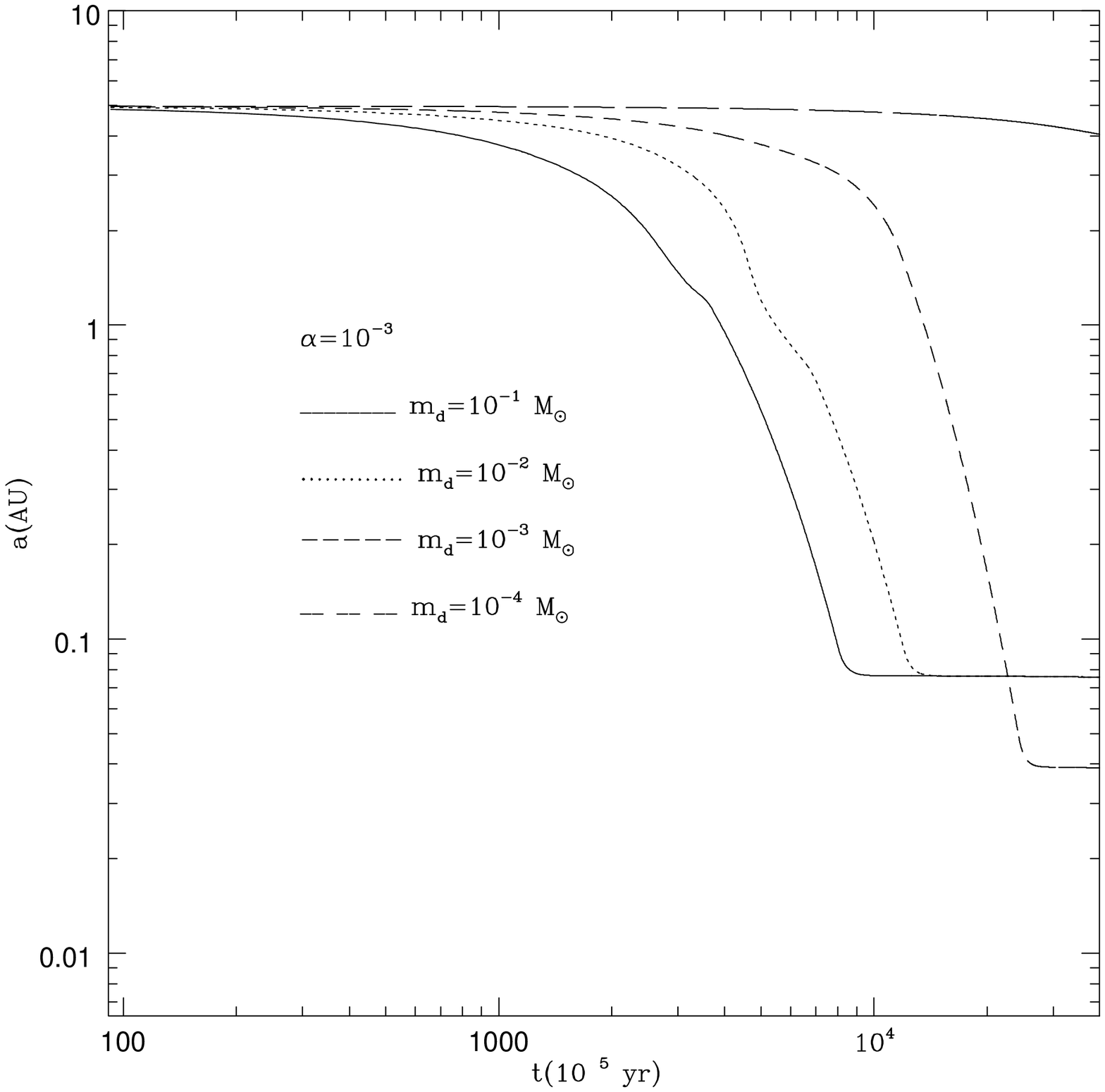,width=8cm} (d)
\vspace{1cm}
\psfig{figure=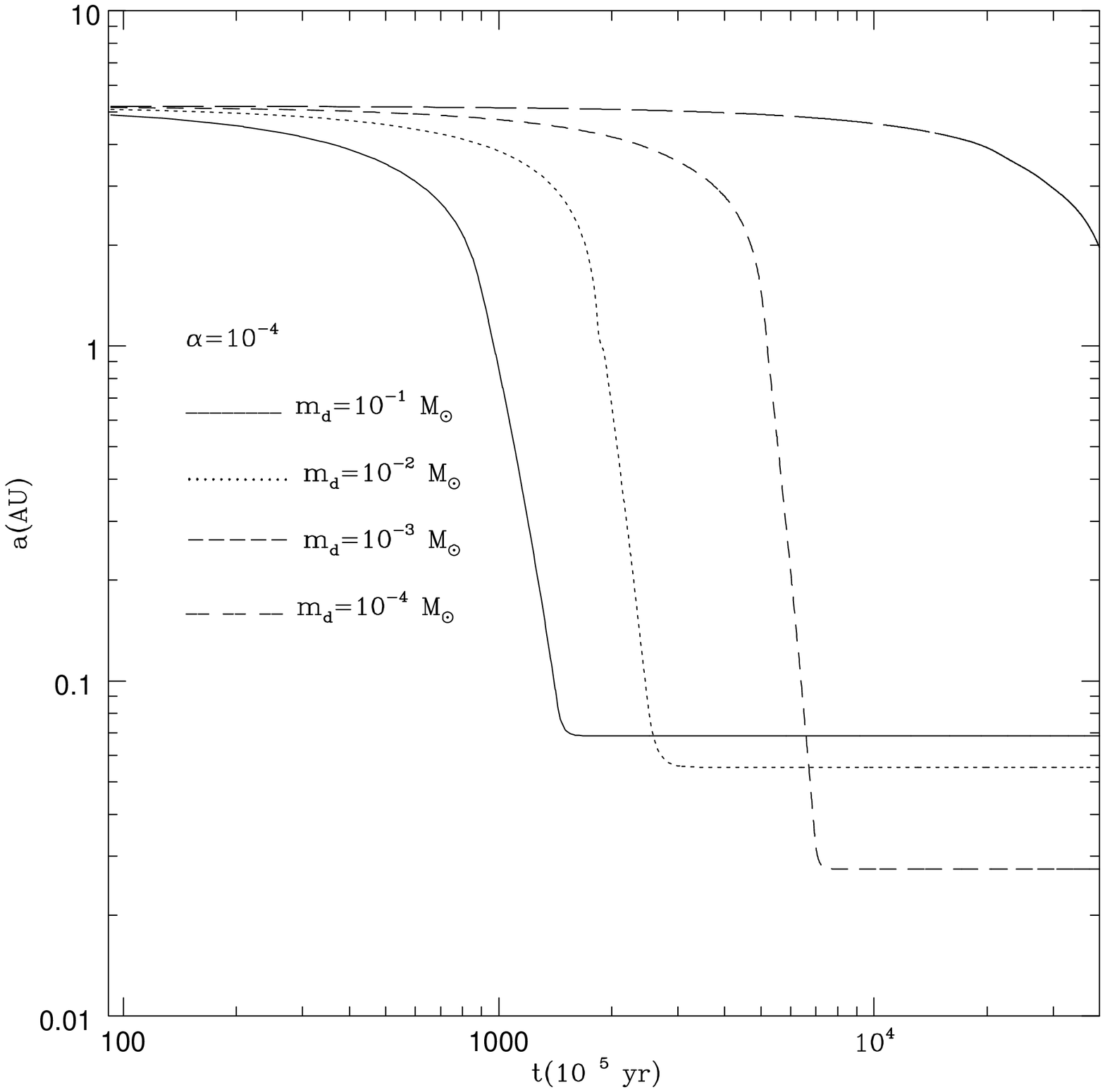,width=8cm}
}}
\caption[]{(a) The evolution of the semi-major axis $a(t)$ of a
Jupiter-mass planet, $M=1 M_{\rm J}$, in a planetesimal disk for
$\alpha=0.1$ and several values of $M_{\rm d}$, $0.1$ (solid line),
$0.01$ (dotted line), $0.001$ (short-dashed line) and $0.0001
M_{\odot}$ (long-dashed line). (b) Same as Fig.6a but with
$\alpha=0.01$. (c) Same as Fig.6a but with $\alpha=0.001$. (d) Same
as Fig.6a but with $\alpha=0.0001$.}
\label{Figmig1}
\end{figure*}

Now, using the converged radial distribution of planetesimals
derived in the previous section, we display in Fig.\ref{Figmig1} the
evolution of the semi-major axis $a(t)$ of a 1 $M_{J}$ planet in such
a disk for $\alpha=0.1$ and $M_{\rm d}=0.1$ (solid line), $0.01$ (dotted
line), $0.001$ (short-dashed line), $0.0001 M_{\odot}$ (long-dashed
line), respectively. We recall here that the planet is initially
located at $5.2$ AU with $i_{\rm p} \sim e_{\rm p} \sim 0$.

We clearly see that for a fixed value of $\alpha$ a disk of larger
mass leads to a more rapid migration of the planet. This behavior is
quite natural since a more massive disk obviously yields a stronger
frictional drag. Thus, in the cases $M_{\rm d}=0.1, 0.01 M_{\odot}$ the planet
migrates to $\simeq 0.08$ AU in $\simeq 1.5 \times 10^9$ yr and to
$\simeq 0.03$ AU in $\simeq 2.5 \times 10^9$ yr, respectively. When
the planet arrives at this distance the dynamical friction switches
off and its migration stops. The stopping is simply due to the inner
radius of the planetesimal disk. The latter is set by the evaporation
radius. Indeed, solid bodies cannot condense at such small orbital
radii $r \la 0.1$ AU because the temperature is too high. Of course,
this evaporation radius $r_{\rm evap}$ depends on the properties of
the solid grains we consider. For instance, for ice particles we have
$r_{\rm evap} \sim 1$ AU (e.g., \citealt{SV2}; \citealt{Kornet1}). In
this article we wish to understand the small orbital radii of observed
planets, over the range $0.03-0.15$ AU. Therefore, we are interested
in the inner regions of the disk where only high-temperature silicates
with $T_{\rm evap} \sim 1350$ K survive. This is why we selected this
component in this study. Then, the main point of Fig.\ref{Figmig1} is
that the properties of solids known to exist in protoplanetary
systems, together with reasonable density profiles for the disk, lead
in a natural fashion to a characteristic radius in the range
$0.03-0.2$ AU for the final semi-major axis of the giant planet. Note
in particular that this process naturally explains why the migration
stops at such radii.

For less massive disks, $M_{\rm d}=10^{-3}, 10^{-4} M_{\odot}$, the
migration is slower and the planet has not enough time to migrate
below $\simeq 4$ AU. Similarly to DP1 and DP2,
note that we plotted 
%stop 
our calculation till $4.5
\times 10^9$ yr which is the typical age of protoplanetary disks like
ours (the Sun is $\sim 4.5 \times 10^9$ yr old). Moreover, the
planetesimal disk should have cleared out by this time (\citealt{ida44}; DP1). 
It is interesting to note that the planet moves closer to the central star
in the case $M_{\rm d}=0.01 M_{\odot}$ than in the case $M_{\rm d}=0.1
M_{\odot}$. This is due to the fact that less massive disks usually
have a smaller surface density which leads to a smaller
temperature. This in turn implies a smaller evaporation radius so that
the planet can move closer to the star.
\footnote{Indeed, we have the energy balance: $T_{\rm e}^4 \propto
\Sigma \nu_{\rm t} \Omega_k^2$ where $T_{\rm e}$ is the effective
temperature, $\Sigma$ the gas surface density, $\nu_{\rm t}$ the
turbulent viscosity and $\Omega_k$ the Keplerian angular velocity.
Besides, the mid-plane temperature $T_c$ obeys the relation $T_c^4
\sim \tau T_{\rm e}^4$ where $\tau$ is the opacity. On the other hand,
the opacity $\tau$ is given by $\tau \sim \kappa \Sigma$ where
$\kappa$ is the Rosseland opacity, while the turbulent viscosity
scales as $\nu_{\rm t} \sim \alpha T_c/\Omega_k$, hence we obtain:
$T_c^3 \propto \kappa \alpha \Sigma^2$.}
In order to study the effect of viscosity on migration, we performed
three other calculations with $\alpha=10^{-2}, 10^{-3}$ and $10^{-4}$,
also plotted in Fig.\ref{Figmig1}. We can note that the dependence of
the final radius on $\alpha$ is rather weak and non-monotonic because
it competes with the dependence on the surface density whose precise
value is an intricate function of $\alpha$. On the other hand, the
migration time usually increases going from $\alpha=10^{-4}$ up to
$\alpha=0.1$. Indeed, as seen from Fig.\ref{FigM1solid} and
Fig.\ref{FigM4solid} a smaller $\alpha$ can lead to a larger mid-plane
solid density. This is partly due to the fact that the dust disk
height is smaller because the turbulence measured by $\alpha$ is
weaker.

The results displayed in Fig.\ref{Figmig1} show that for $\alpha \la
0.01$ a Jupiter-like planet can migrate to a very small distance from
the parent star, $0.03 {\rm AU}<r<0.1 {\rm AU}$, provided the disk
mass is sufficiently large $M_{\rm d} \ga 10^{-3} M_{\odot}$. Only in the
cases $M_{\rm d} \la 10^{-4} M_{\odot}$, or $M_{\rm d} \la 10^{-3} M_{\odot}$ with
$\alpha \ga 0.1$, the interaction with the planetesimal disk is too
weak to yield a significant migration. Note however that a giant
Jupiter-like planet already makes a mass of order $10^{-3}
M_{\odot}$. Hence such objects should arise from protoplanetary disks
with a mass of the order of or larger than $10^{-3} M_{\odot}$. That
is the mass of the initial protoplanetary disk should be at least of
the same order since we do not expect all the matter to end up within
this giant planet during its formation stage.
\footnote{However, if the giant planet forms through a different
process, e.g., the instability of a distinct cloud, which has no
relation with the disk itself, then the latter may have any mass.}
Then, our results show that a final radius of $0.03 {\rm AU}<r<0.1
{\rm AU}$ is a natural outcome, provided the planetesimal disk has not
been cleared off too early on (e.g., by gravitational scattering).

Summarizing, the present model predicts that, unless the disk mass is
very small $M_{\rm d} \la 10^{-4} M_{\odot}$, planets tend to move close to
the parent star and to pile up to distances of the order of
$0.03-0.04$ AU. However, with some degree of fine-tuning it is also
possible to find a planet at intermediate distances between its
formation site and such small radii.

Before concluding this section, we have to add some hints of the way migration was calculated 
in the case the disc mass, $M_{\rm d}$, is smaller than that of the protoplanet. 
One can think that less massive discs are not able to move a more massive planet. This idea 
is not completely correct, as we shall see.
As we know, when the planet mass is less than or comparable to the local disc mass,  
the planet behaves as a representative particle in the disc, as shown by Lin \& Papaloizou (1986).
When the protoplanet has a mass larger than that of the disc the satellite acts as a dam against the 
viscous evolution of the disc, and can lead to a substantial change in the disc structure in the 
vicinity of the planet. The coupled disc-planet evolution in this case has been studied by Syer \& Clarke (1995) and 
by Ivanov et al. (1999), who showed that the inertia of the planet plays an important role in this case (see Nelson et al. 1999, Eq. 9). The effect of the interaction between the secondary and the disc leads to accumulation of the disc matter 
in the region behind the protoplanet (see Eq. 12 and 39 of Syer \& Clarke 1995). In order to calculate the migration, 
in the quoted case, we used the surface density given in Ivanov et al. (1999) in the model of migration described in the previous sections. 

%
%%A comparison of the previous results with that of DP2 shows several differences. Starting with the case $\alpha=10^{-3}$, %%we see that in all mass cases migration is more rapid in DP2 with respect to the prediction of this paper, due to the %%different disc model adopted. 
%

\subsection{Distribution of orbital periods}

\begin{figure*}
\centerline{\hbox{(a)
\psfig{figure=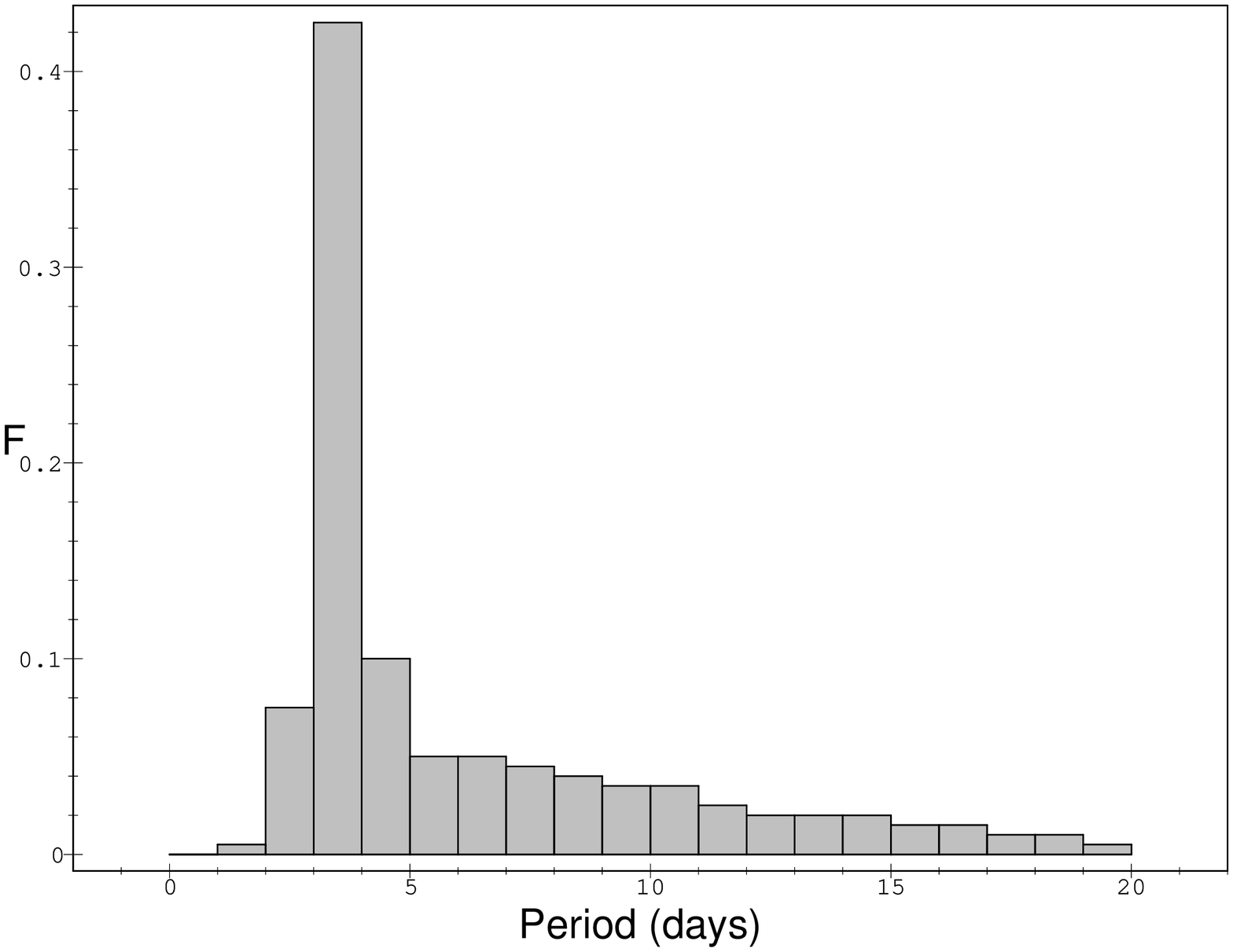,width=6.3cm,height=7.6cm,angle=270} (b)
\psfig{figure=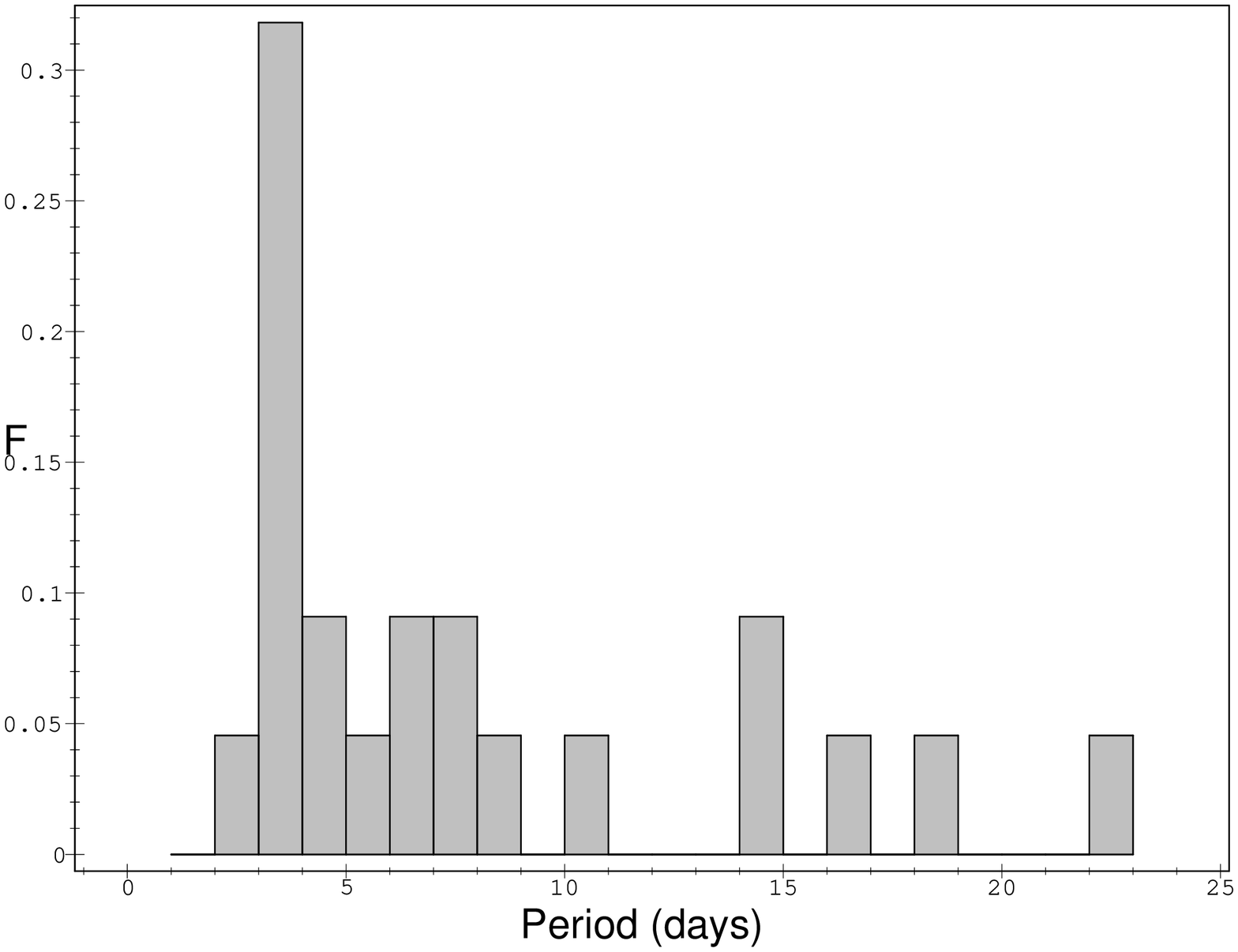,width=6.3cm,height=7.6cm,angle=270}
}}
\caption[]{(a) Fraction of planets having orbital periods in the range
0-20 days, calculated according the model of this paper.  (b) Fraction
of planets having orbital periods in the range 0-20 days, calculated
according to data (see www.exoplanets.org).}
\label{Figperiod1}
\end{figure*}

Finally, we show in Fig.\ref{Figperiod1} the predictions of our model
for the distribution of planets in the inner part of the disk. More
precisely, we plot the fraction of planets in the orbital period range
0-20 days calculated from the model described in the previous section,
assuming a uniform probability distribution in the plane
$(\log(\alpha),\log(M_{\rm d}))$ in the range $10^{-4} < \alpha < 10^{-1}$
and $10^{-4} < M_{\rm d} < 10^{-1} M_{\odot}$. This is a rather arbitrary
choice but our point is simply to check whether the observed
distribution can be explained by reasonable values for $\alpha$ and
$M_{\rm d}$ or whether it requires some fine tuning. The right panel
represents the same distribution obtained with the data given in
www.exoplanets.org, see also \citet{Kuchner1}. We can see that we
actually obtain a reasonable agreement with the data. Of course, there
is some degeneracy so that different probability distributions in the
plane $(\log(\alpha),\log(M_{\rm d}))$ could yield the same results. However,
the main point of Fig.\ref{Figperiod1} is to show that the observed
distribution of orbital periods can be easily recovered from our
model, without any fine-tuning and with reasonable values for $\alpha$
and $M_{\rm d}$. Moreover, as can be seen from Fig.\ref{Figmig1}, within our
model the peak at 3-4 days for the orbital period comes from disks
with $M_{\rm d} \ga 10^{-3} M_{\odot}$ and $\alpha \la 10^{-2}$. As noticed
in the previous section this result actually fits nicely with the data
since we can expect Jupiter-like planets to form in such
protoplanetary disks with $M_{\rm d} \ga 10^{-3} M_{\odot}$ (note also that
the dependence on $\alpha$ is rather weak).

Of the $\simeq 20$ planets with periods less than 20 days, 7 have
periods in the range 3-4 days, and it seems that this trend is not an
artifact of observational selection (see \citealt{Kuchner1}). As
stressed by \citet{Kuchner1}, phenomena like the interactions among
two planets and a star, that can leave a planet trapped by stellar
tides into a circular orbit of $\simeq 0.04$ AU (\citealt{Rasio1}), or
halting of planet migration by loss of mass to the star
(\citealt{Trilling1}), are rare. The migration through resonant
interaction with planetesimals in the disk (\citealt{Murray1})
provides a natural way of halting the planet but unfortunately to
change the orbit of a Jupiter-mass planet requires approximatively a
Jupiter mass of planetesimals and a too massive disk. The other
possibility to explain the observed distribution of planets seen in
Fig.\ref{Figperiod1} is connected with a gas disk truncated at a
temperature of 1500 ${\rm K}$ by the onset of Magneto-Rotational
instability (\citealt{Kuchner1}).  In other terms, disk temperature
determines the orbital radii of the innermost surviving planets.  In
our model, the distribution of internal planets is naturally connected
to the evaporation of planetesimals at very short radii. As stressed
in DP1, DP2, this model has not the drawback of \citet{Murray1}, namely
that of requiring a too large disk mass for migration and at the same
time it has the advantage of \citet{Murray1} of having an intrinsic
natural mechanism that provides halting of migration.

%1) Parlare della relazione posizione pianeti, modello? 
%(parte precedente tolta)

%2) Accenno metallicità?

\section{Conclusions}

In this paper, we further developed the model for the migration of
planets introduced in DP1 and extended to time-dependent planetesimal
accretion disks in DP2. After releasing the assumption of DP2 that the
surface density of planetesimals is proportional to that of gas, we
used a simplified model developed by Stepinski \& Valageas (1996,
1997), that is able to simultaneously follow the evolution of gas and
high-temperature silicates for up to $10^7 {\rm yr}$. Then we coupled
this disk model to the migration model introduced in DP1 in order to
obtain the migration rate of the planets in the planetesimal disk and
to study how the migration rate depends on the disk mass, on its time
evolution and on the dimensionless viscosity parameter $\alpha$.

We found that in the case of disks having a total mass of $M_{\rm d} >
10^{-3} M_{\odot}$ planets can migrate inward over a large distance,
while if $M_{\rm d} < 10^{-3} M_{\odot}$ the planets remain almost at their
initial position. On the other hand, for $M_{\rm d} \sim 10^{-3} M_{\odot}$
a significant migration requires $\alpha \la 10^{-2}$.

If the migration is efficient the planet usually ends up at a small
radius in the range $0.03 - 0.1$ AU which is simply set by the
evaporation radius of the gaseous disk which gave rise at earlier
times to the radial distribution of the planetesimal swarm. Thus, our
model provides a natural explanation for the small observed radii of
extra-solar giant planets. In particular, the halting of the inward
migration of the planet is intrinsic to this process and it does not
require a second mechanism.

Finally, we noticed that the observed distribution of planet periods
in the range 0-20 days can be easily obtained within this framework
without fine-tuning. In particular, such small radii naturally occur
for Jupiter-like planets if the initial disk has a mass of the same
order or larger, which is quite likely. In order to inhibit this
process (so that Jupiter-like planets like our own remain at larger
distances $\ga 5$ AU) the planetesimal disk must be cleared off over a
time-scale of the order of or smaller than $10^9$ yr (depending on the
properties of the disk) or the disk mass must be rather small
(i.e. smaller than $1 M_{\rm J}$) which could suggest an alternative
formation scenario for such a giant planet (i.e. not related with the
disk itself).

\section*{Acknowledgments}
%%%%%%Work partially supported by funds ex-60\% 98.
%%V.A.-D. would like to thank Prof. Giuseppe Moncada.
We are grateful to E. N. Ercan for stimulating discussions during
the period in which this work was performed. A. Del Popolo would
like to thank ASI and 
Bo$\breve{g}azi$\c{c}i University Research
Foundation and for the financial support through the project code
01B304.

\end{document}